\documentclass[aps,prb,twocolumn,notitlepage,longbibliography,showpacs,floatfix]{revtex4-2}

\usepackage{graphicx}
\usepackage{amsmath,amssymb,bbold,bm,color}
\usepackage{siunitx}
\usepackage{float}
\usepackage{bbm}
\usepackage{epstopdf}
\usepackage{hyperref}
\usepackage{booktabs}
\usepackage{comment}
\usepackage[dvipsnames]{xcolor}
\usepackage[normalem]{ulem}

\usepackage[resetlabels,labeled]{multibib}

\newcites{New}{The other list}

\newcommand{\bk}{{\bm k}}

\newcommand{\br}{{\bm r}}

\newcommand{\bR}{{\bm R}}

\newcommand{\cT}{{\cal T}}

\newcommand{\cH}{{\cal H}}
\newcommand{\cG}{{\cal G}}

\newcommand{\bee}{\begin{equation}}
\newcommand{\ee}{\end{equation}}

\hypersetup{colorlinks=true,linkcolor=blue,citecolor=blue,urlcolor=blue}

\begin{document}
\title{Edge currents as probe of topology in twisted cuprate bilayers}
\author{Vedangi Pathak} \thanks{Corresponding author: vedangi@phas.ubc.ca.} 
\affiliation{Department of Physics and Astronomy \& Stewart Blusson Quantum Matter Institute,
University of British Columbia, Vancouver BC, Canada V6T 1Z4}
\author{Oguzhan Can} 
\affiliation{Department of Physics and Astronomy \& Stewart Blusson Quantum Matter Institute,
University of British Columbia, Vancouver BC, Canada V6T 1Z4}
\author{Marcel Franz} 
\affiliation{Department of Physics and Astronomy \& Stewart Blusson Quantum Matter Institute,
University of British Columbia, Vancouver BC, Canada V6T 1Z4}
\date{\today}
\begin{abstract}
Bilayers made of high-$T_c$ cuprate superconductor Bi$_2$Sr$_2$CaCu$_2$O$_{8+x}$ assembled with a twist angle close to $45^\circ$ have been recently shown to spontaneously break time reversal symmetry $\cT$, consistent with theoretical predictions for emergent chiral topological $d_{x^2-y^2}+id_{xy}$ phase in such twisted $d$-wave superconductors. Here we use a minimal microscopic model to 
estimate the size of spontaneous chiral edge currents expected to occur in the $\cT$-broken phase. In accord with previous theoretical studies of chiral $d$-wave superconductors we find small but non-vanishing edge currents whose magnitude and spatial distribution are sensitive to the type of the edge or domain wall. We nevertheless predict these to be above the detection threshold of the state-of-the-art magnetic scanning probe microscopy. In addition, by deriving a simple relation between the edge current and the electron spectral function we help elucidate the longstanding disparity between the size of edge currents in chiral $d$-wave and $p$-wave superconductors.

\end{abstract}

\maketitle

\section{Introduction}
Chiral superconductors are characterized by the order parameter $\Delta_k\simeq\Delta_0((k_x+ik_y)/k_F)^m$ with integer $m>0$ being both the magnetic quantum number which determines the orbital angular momentum (OAM) $m\hbar$ per Cooper pair and the Chern number which is a topological invariant characterizing the Bogoliubov-de Gennes (BdG) spectrum and determining number of the chiral edge modes \cite{Stone2004,Huang2014,Kallin2016}. Such an order parameter breaks time reversal symmetry allowing spontaneous currents which we expect to find at the edges of the sample. The edge modes can support the supercurrent, although some states extend into the bulk may also participate \cite{Stone2004}. Unlike the Chern insulator, the edge currents in chiral superconductors are not quantized \cite{Huang2015}. The reason is that these states are mixtures of particle and hole degrees of freedom and how much electrical current they carry does not only depend on the dispersion but also on the particle-hole content of the wavefunctions. Therefore we might not find current where we expect due to cancellations between particle and hole degrees of freedom \cite{Huang2014}. The edge modes give rise to quantized thermal conductance which is, however, notoriously difficult to measure. 

Edge currents predicted for a chiral $p$-wave superconductor ($m=1$) within the quasi-classical approximation using a continuum model \cite{Stone2004} are consistent with OAM $L_z=N\hbar/2$ for $N$ fermions for a disk geometry. This quasi-classical analysis was generalized in Ref.\ \cite{Huang2014} to higher angular momentum pairing channels ($m>1$) and it was found that edge currents vanish for single band continuum models. This is so even though in the strong coupling (BEC) limit the expected OAM is $L_z=Nm\hbar/2$; vanishing edge currents in the simplest model suggests that this is not the case. This reduction of angular momentum for higher values of chirality was attributed to spectral flow along the edge states in Ref.\ \cite{Tada2015}.

Edge currents for higher chirality states have not received as much attention as the chiral $p$-wave state (long thought relevant to a candidate triplet superconductor Sr$_2$RuO$_4$), owing chiefly to the paucity of realistic candidate systems. 
Nevertheless theoretical studies of more complex models \cite{Wang2018,Rainer1998,Horovitz2003,Braunecker2005,Huang2014,Black-Schaffer2012}
have found non-vanishing edge currents in the chiral $d$-wave state. In contrast to the quasi-classical result for single-band continuum theory, Ref.\ \cite{Huang2014,Wang2018} found nonzero edge currents for the $m>1$ case based on self-consistent BdG lattice models of chiral $d$-wave state. Refs. \cite{Suzuki2016, Holmvall2023_2} investigated edge currents in a finite-sized disk geometry, solving quasiclassical Eilenberger equations, and determined that screening does not alter the qualitative behavior of edge currents. However, all these studies concluded that the chiral $d$-wave has edge currents an order of magnitude smaller than what is expected for chiral $p$-wave.  
 
Here we consider a specific 
candidate for the chiral $d$-wave superconducting state realized in the 2D heterostructure of twisted bilayer cuprates such as Bi$_2$Sr$_2$CaCu$_2$O$_{8+x}$ (Bi-2212). Composed of two monolayer $d$-wave superconductors ($d$SC) stacked with a near $45^\circ$ twist such a structure is predicted to host a topological $d_{x^2-y^2}+id_{xy}$ state (abbreviated henceforth as $d+id'$) with bulk gap and chiral edge modes  \cite{Can2021}. Spontaneous $\cT$-breaking here occurs due to the symmetry-imposed vanishing of the first harmonic in the Josephson energy between the two layers at $45^\circ$ twist, with the second harmonic favoring the $\pi/2$ phase difference. Recent transport experiments on twisted Bi-2212 flakes indeed reported fractional Shapiro steps and Fraunhofer patterns consistent with strong second harmonic Josephson energy \cite{Zhao2021}. The same study also observed a pronounced zero-field superconducting diode effect in near-$45^\circ$ twisted samples, 
indicating spontaneous $\cT$-breaking at the interface \cite{Etienne2023}. Other transport experiments, however, reported only conventional behavior in twisted Bi-2212 junctions  \cite{Xue2021} and more recently also in twisted Bi$_2$Sr$_2$CaCuO$_{6+x}$ (Bi-2201) \cite{Xue2023}. The most recent transport study of twisted Bi-2212 junctions reported \cite{Martini2024}, once again, an unconventional behavior consistent with theoretical predictions for twisted $d$-wave superconductors.

On the theory side, more detailed microscopic studies of twisted cuprate bilayers that incorporate the effect of strong correlations \cite{Song2022,Lu2022,Spalek2023,Pixley2023,Senechal2024a,Senechal2024b}, inhomogeneity \cite{Haenel2022,Yuan2023} and the effect of applied current and magnetic field \cite{Volkov2023} make predictions for the $\cT$-broken phase that differ in important details such as the critical twist angle and the size of the induced excitation gap.  
Given this situation it is important to explore the full range of physical manifestations of the predicted chiral $d+id'$ phase.  In addition to the fractional Shapiro steps, Fraunhofer patterns and the diode effect mentioned above signatures of $\cT$-breaking can be probed directly via polar Kerr effect measurements as proposed in Ref.\ \cite{Can2021_2}. Fractional and coreless vortices have been predicted to occur in the chiral $d$SC \cite{Holmvall2023}, which could be experimentally detected.    

Perhaps the most persuasive direct manifestation of the chiral $d+id'$ state, however, would be observation of the spontaneous edge currents associated with the topologically protected edge modes. The aim of this work is to evaluate the edge currents resulting from the spontaneous $\cT$ breaking in a lattice model of a $d+id'$ superconducting system designed to capture the physics of twisted Bi-2212 bilayer. To this end we first construct a suitable `aligned lattice model' and solve the resulting Bogoliubov - de Gennes (BdG) theory self-consistently in the long-strip geometry with several types of edges and domain walls that produce spontaneous currents. Based on this solution we calculate the edge currents and the resulting magnetic fields which we find to be strong enough to be detectable with a high-resolution scanning SQUID microscope.  As appropriate for a 2D sample of a strongly type-II cuprate superconductor our treatment throughout the paper ignores the Meissner screening.

It is important to note that for a pure, single-layer $d$SC, a different $\cT$ broken state has been predicted to occur near a pair-breaking boundary, e.g.\ one formed by a (110) edge in a $d_{x^2-y^2}$ superconductor \cite{Hakansson2015,Holmvall2018,Holmvall2019,Wennerdal2020}.
In such a state fractional vortex anti-vortex pairs are predicted to spontaneously nucleate along the boundary below $T^*\approx 0.18 T_c$, breaking also the translation symmetry along the edge. In our self-consistent treatment of a single-layer $d_{x^2-y^2}$ superconductor 
with a (110) edge, we were able to confirm the onset of this $\cT$ broken phase below $T^*$. Nevertheless this effect remains experimentally unconfirmed \cite{Suzuki2016} and we defer a study of its interplay with the bulk $\cT$ broken phase to future work.

In the remainder of this paper we focus on types of edges and domain walls that are not pair-breaking in the above sense and thus allow us to study edge currents intrinsic to the bulk $\cT$-broken chiral $d+id'$ phase that emerges in the bilayer near a $45^\circ$ twist.
In addition, we focus on bilayer solutions that retain translation invariance along all edges.   

\begin{figure}
    \centering
    \includegraphics[width=\linewidth]{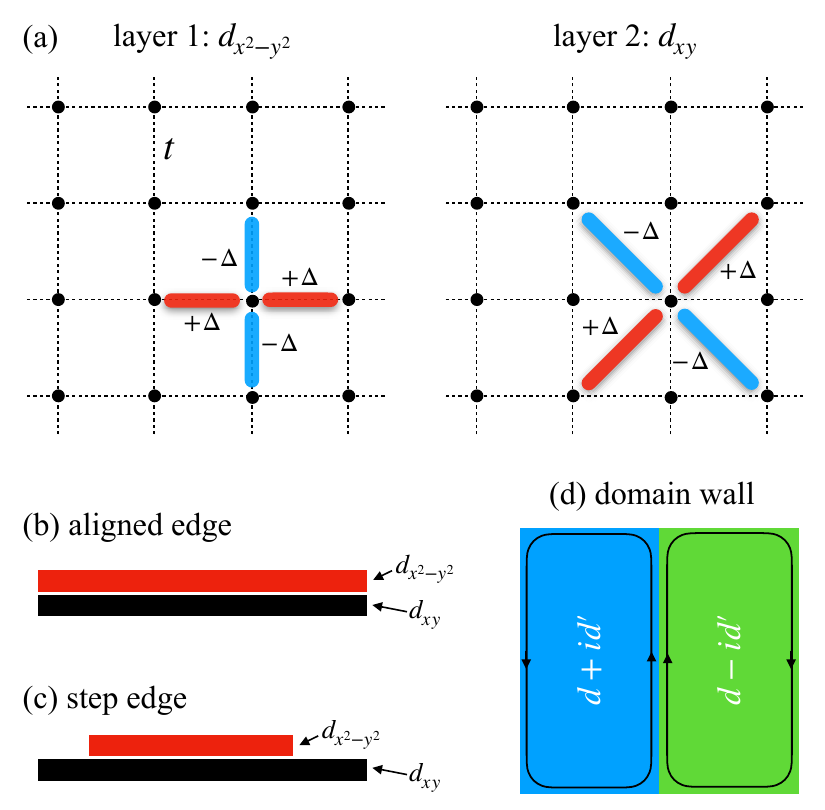}
    \caption{(a) Real-space representation of the aligned lattice model for the $45^\circ$ twisted bilayer. Electrons in both layers move on identical square lattices. The twist is implemented by adopting attractive interactions that favor a $d_{x^2-y^2}$ ($d_{xy}$) order parameter in layers 1 (2), respectively. The extended $s$-wave order parameters have the same spatial structure except the bond fields do not alternate sign. A side view of (b) aligned and (c) step edge. Panel (d) shows a domain wall between $d+id'$ and $d-id'$ phases. The arrows indicate edge currents and illustrate why we expect a domain wall to carry twice the current of an edge. }
    \label{fig:0}
\end{figure}

\section{Modelling the twisted bilayer}

The bulk behavior of a twisted cuprate bilayer is most conveniently described by a continuum BCS theory which has been extensively checked against microscopic lattice models with the correct Fermi surface structure \cite{Can2021,Can2021_2,Tummuru2022}. Because we wish to model behavior of the twisted bilayer near an edge it is preferable to use a lattice model. However, simulating the full microscopic lattice model near the $45^\circ$ twist tends to be computationally expensive owing to its large moir\'e unit cell.  To avoid complications arising from such a large unit cell we employ here the `aligned lattice' model introduced in Ref.\ \cite{Can2021_2}. The idea is to represent two monolayers by two perfectly aligned square lattices. The $45^\circ$ twist is then implemented by imposing a $d_{x^2-y^2}$ order parameter in one layer and a $d_{xy}$ order parameter in the other, as illustrated in Fig.\ \ref{fig:0}(a). 

It is important to bear in mind that by keeping the unit cell small the aligned lattice model necessarily misses all the physics related to the Brillouin zone folding due to the formation of the moir\'e lattice. On the other hand, as verified in Refs.\ \cite{Can2021,Can2021_2,Tummuru2022}, the model still supports the correct phenomenology as compared with the full microscopic calculation performed with the large moir\'e unit cell, the continuum BdG theory as well as the Ginzburg-Landau approach.

\subsection{The aligned lattice model}

The Hamiltonian of the aligned lattice model is given by $H=H_0+H_{\rm int}$ with
\begin{align}\label{eq:h1}
H_0=&-t\sum_{\langle ij\rangle,\sigma a} (c^\dagger_{i\sigma a}c_{j\sigma a}+{\rm h.c.})
-\mu\sum_{i,\sigma a} n_{i\sigma a} \\
&-g\sum_{i,\sigma}(c^\dagger_{i\sigma 1}c_{i\sigma 2}+{\rm h.c.}),\nonumber
\end{align}
describing the normal-state tight-binding band structure of the bilayer. Here $c^\dagger_{i\sigma a}$ creates an electron on site $i$ with spin $\sigma$ in layer $a=1,2$, $\langle ij\rangle$ denotes summation over nearest neighbor sites and $n_{i\sigma a}=c^\dagger_{i\sigma a}c_{i\sigma a}$ is the number operator.  $t$ and $g$ denote respectively the in-plane and interplane tunneling amplitudes and $\mu$ is the chemical potential.

$H_{\rm int}$ describes attractive electron-electron interactions that give rise to $d$-wave superconductivity in the individual layers, 
\begin{align}
H_{\rm int}=&-V_1\sum_{\langle ij\rangle,\sigma\sigma'} n_{i\sigma 1}n_{j\sigma' 1}
-V_2\sum_{\langle\langle ij\rangle\rangle,\sigma\sigma' } n_{i\sigma 2}n_{j\sigma' 2} \label{eq:h2}
\end{align}
where $\langle\langle ij\rangle\rangle$ denotes summation over second-neighbor sites on the square lattice. For positive $V_a$ (and assuming decoupled layers) this form of interaction is known to produce a $d_{x^2-y^2}$ order in layer 1 and $d_{xy}$ order in layer 2.    

Performing the standard mean-field decoupling of the interaction term \eqref{eq:h2} in the pairing channel one obtains the BdG Hamiltonian. For a uniform system with periodic boundary conditions, it can be written compactly in the momentum space as  $\mathcal{H}=\sum_\bk\Psi^\dag_\bk h_\bk \Psi_\bk$ where $\Psi_\bk=(c_{\bk\uparrow 1},c^\dag_{-\bk\downarrow 1}, c_{\bk\uparrow 2},c^\dag_{-\bk\downarrow 2})^{T}$ and 
\begin{equation}\label{eq:h4}
  h_\bk=
 \begin{pmatrix}
   \xi_\bk & \Delta_{\bk 1} & g & 0 \\
   \Delta_{\bk 1} ^\ast & -\xi_\bk & 0 & -g \\
   g & 0 &   \xi_\bk & \Delta_{\bk 2} \\
   0 & -g &    \Delta_{\bk 2} ^\ast & -\xi_\bk
  \end{pmatrix}.
\end{equation}
 The normal state for each monolayer has a dispersion $\xi_\bk = -2t(\cos{k_x}  + \cos{k_y}) - \mu$. The superconducting order parameters assume the form  
\begin{align}\label{eq:d_sc}      
\Delta_{\bk 1}&=\Delta_{1d}(\cos k_x - \cos k_y), \nonumber \\
\Delta_{\bk 2}&=\Delta_{2d}(2\sin k_x\sin k_y),
\end{align} 
and their amplitudes are determined self-consistently from the 
gap equation,
\begin{align}\label{eq:self-consistency}
    \Delta_{ad}&=2{V_a}\sum_{\mathbf{k},\alpha}\frac{\partial E_{\mathbf{k}\alpha}}{\partial \Delta_{ad}^*}\tanh{\left(\frac{\beta E_{\mathbf{k}\alpha}}{2}\right)}\nonumber\\
   &=2{V_a }\sum_{\mathbf{k},\alpha}\langle\bk\alpha|\frac{\partial h_{\mathbf{k}}}{\partial\Delta_{ad}^*}|\bk\alpha\rangle \tanh{\left(\frac{\beta E_{\mathbf{k}\alpha}}{2}\right)},
\end{align}
where $\beta=1/k_BT$ is the inverse temperature, $|\bk\alpha\rangle$ are eigenstates of $h_\bk$ belonging to positive eigenvalues $E_{\bk\alpha}$ $(\alpha=1,2)$, and the second line is convenient in numerical computations. 

We find that for decoupled monolayers a $d$SC with non-zero order parameters defined in Eq.\ \eqref{eq:d_sc} represents a stable ground state of the BdG theory for much of the phase diagram spanned by parameters $V_1$, $V_2$ and $\mu$. However, a state with extended $s$-wave order parameters defined as 
\begin{align}\label{eq:s_sc}
    \Delta_{\bk 1,s}&=\Delta_{1s}(\cos k_x + \cos k_y), \nonumber \\
    \Delta_{\bk 2,s}&=\Delta_{2s}(2\cos k_x\cos k_y),
\end{align}
is a close competitor. Here $\Delta_{as}$ are given by equations analogous to Eqs.\ \eqref{eq:self-consistency}. As discussed in more detail below this competing $s$-wave state tends to crop up in situations where the dominant $d$-wave order parameter is spatially varying, e.g.\ near the edges.

\subsection{Model parameters}

\begin{figure}
    \centering    \includegraphics[width=0.8\linewidth]{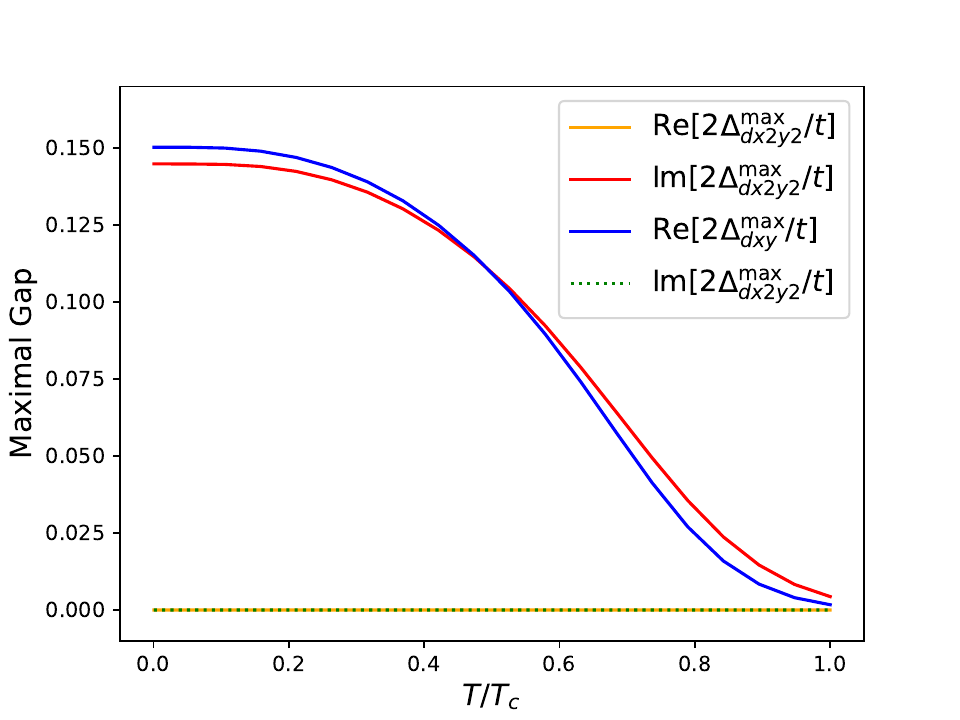}
    \caption[Self-consistent solution of the superconducting order parameters in the bulk stabilizing a $d+id'$ bilayer as a function of temperature for the long strip geometry with the aligned edge configuration and the step edge configuration.]{Self-consistent solution of the maximal gap from superconducting order parameters in the bulk stabilizing a $d+id'$ bilayer as a function of temperature. A $d_{x^2-y^2}$ and a $d_{xy}$ emerge with a relative phase of $\pi/2$ between them. The data is shown for the continuum Hamiltonian for a weak inter-layer coupling $g=8$~meV.}
    \label{fig:sc_bulk}
\end{figure}

We set our normal state parameters as $t=0.38$~eV, $\mu=-1.2t$ and the interaction strengths as $V_1\approx0.26$~eV, $V_2\approx0.23$~eV. 
These parameters give us a near-circular Fermi surface, a $d$-wave superconducting order parameter, and a maximal gap of approximately 55~meV in each monolayer, in accord with the observed gap size in optimally doped Bi-2212. The calculation of maximal gap at Fermi-surface is provided in Appendix~\ref{app:max_gap}. 
We show the temperature dependence of the maximal gap for the bulk d+id' superconductor in Fig.~\ref{fig:sc_bulk}. We use these parameters for all subsequent calculations except where noted otherwise. 

Upon weakly coupling the layers, a self-consistent solution develops a relative phase of $\pi/2$ between the superconducting order parameters, thereby reproducing the expected $d+id'$ order parameter in the twisted bilayer system. 
Our chosen value of the interlayer coupling $g$ follows from the estimate for twisted Bi-2212 bilayers given in Ref.\ \cite{Tummuru2022} based on experimental data of Ref.\ \cite{Zhao2021}. It is to be noted, however, that estimates of $g$ vary widely in the available literature. Also, we find that the edge current is quite sensitive to the value of $g$, with larger values generally supporting stronger currents. For this reason, we shall give results for edge currents and the associated magnetic fields for several representative values of $g=4-19$ meV covering the range most likely relevant to Bi-2212.    

We note here that the BdG theory yields physical, current-conserving solutions only when the order parameters are determined self-consistently.  However, obtaining a fully self-consistent solution when working with various constrained geometries can get numerically expensive. This is precisely the reason why we choose to work with a simple aligned lattice two-band model instead of the full microscopic model describing the large moir\'e unit cell of the bilayer system.  The advantage of a two-band model is that it can accurately reproduce the physical observables while being numerically tractable, especially when working with long strip geometries as described in the following subsection.

\subsection{Long strip geometry}\label{sec:configs}

We use a long strip geometry to examine the characteristic edge effects in this system. In all cases we assume translational invariance along the $x$-direction  and impose periodic boundary conditions in this `long direction'. The strip has a finite width of $N_y$ unit cells along the $y$-direction and the boundary conditions in the $y$-direction are chosen based on the specific  configuration we consider. We study three different configurations of the edges of this system which we discuss below.

Accordingly, we convert the mean-field BdG theory following from the Hamiltonian  Eqs.~(\ref{eq:h1},\ref{eq:h2}) to the strip geometry by taking a partial Fourier transform along $x$ using 
\begin{equation}
 c_{y\sigma a}(k)=\sum_{x}\frac{\text{e}^{ik x}}{\sqrt{N_x}}c_{(x,y)\sigma a}.   
\end{equation}
Here $\bR=(x,y)$ is used to denote lattice site position and $N_x$ is the number of unit cells along the periodic direction.
In this representation the BdG Hamiltonian ${\cal H}(k)$ becomes a matrix of size  
$4N_y$ for each value of the crystal momentum $k$ drawn from a 1D Brillouin zone $(-\pi/a,\pi/a)$. We diagonalize this matrix numerically and from its eigenvectors and eigenvalues compute the order parameters and currentts which now vary spatially along $y$.  

\subsubsection{Aligned edge configuration}

\begin{figure}
    \centering    \includegraphics[width=\linewidth]{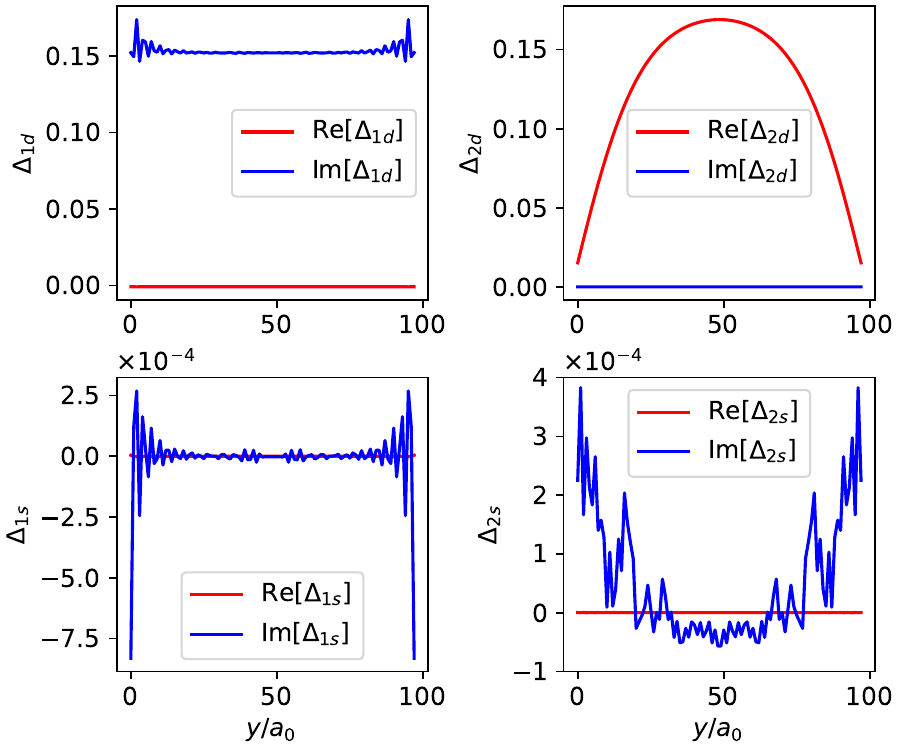}
    \caption[Spatial profile of the superconducting order parameters of the aligned edge configuration in the long strip geometry as a function of $y$.]{Spatial profile of the superconducting order parameters of the aligned edge configuration in the long strip geometry as a function of $y$. The top panel shows the self-consistent solutions of the $d-$wave order parameters in layers 1 and 2 labeled as $\Delta_{1d}$ ($d_{x^2-y^2}$) and $\Delta_{2d}$ ($d_{xy}$) respectively. The bottom panel shows the $s$-wave order parameters in layers 1 and 2, labeled as $\Delta_{1s}$ and $\Delta_{2s}$ respectively. The data depicts maximal superconducting gap at Fermi surface normalized with respect to $t$ for a weak inter-layer coupling and a temperature of $T/T_c\approx0.001$.}
    \label{fig:sc_aligned}
\end{figure}

In this configuration, the edges of layers 1 and 2 are perfectly aligned as shown in Fig.~\ref{fig:0}(b). We simulate this scenario using the long-strip geometry with open boundary conditions in the $y$-direction for both layers. 

Our self-consistent calculations for each monolayer, reveal that considering nearest-neighbor attractive pairing adequately stabilizes a $d$-wave pairing state in the bulk for a strip width $N_y$. This state is sustained in each monolayer, across temperatures up to $T_c$, with the maximum bulk gap of approximately $55$ meV at low temperatures. With the chemical potential set to $\mu=-1.2t$, introducing weak interlayer coupling yields a $d_{x^2-y^2}$ order parameter in layer 1 and a $d_{xy}$ order parameter in layer 2, exhibiting a relative phase of $\pi/2$. 

The spatial profile of the onsite order parameters is shown in Fig.~\ref{fig:sc_aligned}. The edge is pair-breaking for the $d_{xy}$ order parameter in layer 2. This leads to a large suppression of $\Delta_{2d}$ near the edge and a nucleation of an extended $s$-wave order parameter in layer 2. The phase difference between the $s$-wave order parameter and the $d_{xy}$ order parameter is $\pi/2$, giving rise to the $d_{xy}+is$ order parameter in layer 2 near the edge. This is a manifestation of the pair-breaking edge physics explored previously in Ref.\ \cite{Hakansson2015}. This local $\cT$-breaking occurs already for a single-layer $d$SC and is not relevant to our discussion of edge currents in the chiral $d+id'$ phase in twisted bilayer.  
The effect complicates our analysis of the bilayer as the $d+is$ order parameter is time-reversal symmetry breaking and can give rise to edge currents, a topic addressed in Sec.~\ref{sec:currents}. For this reason we focus in the following on types of edges that are not pair breaking and therefore avoid formation of the local $d+is$ phase.  

\subsubsection{Step edge configuration}
\begin{figure}
    \centering
    \includegraphics[width=\linewidth]{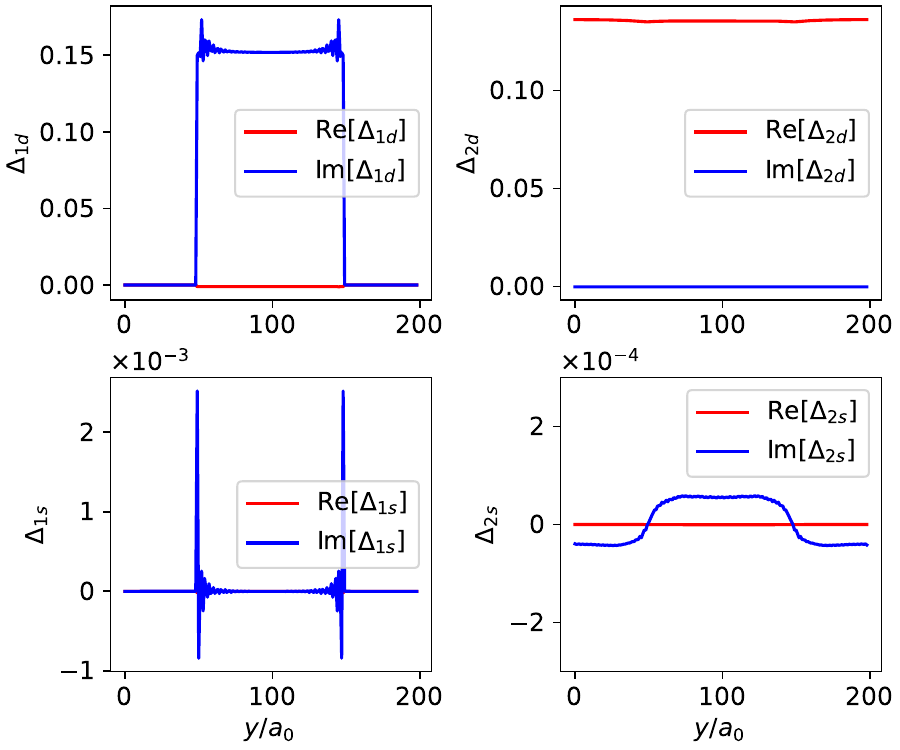}
    \caption[Spatial profile of the superconducting order parameters of the step edge configuration in the long strip geometry as a function of $y$.]{Spatial profile of the superconducting order parameters of the step edge configuration in the long strip geometry as a function of $y$. The top panel shows the self-consistent solutions of the $d-$wave order parameters in layers 1 and 2, labeled as $\Delta_{1d}$ ($d_{x^2-y^2}$) and $\Delta_{2d}$ ($d_{xy}$) respectively.
    The bottom panel shows the $s$-wave order parameters in layers 1 and 2, labeled as $\Delta_{1s}$ and $\Delta_{2s}$ respectively. The data depicts maximal superconducting gap at Fermi surface normalized with respect to $t$ for a weak inter-layer coupling ($g\approx8$~meV) and a temperature of $T/T_c\approx0.001$.}
    \label{fig:sc_step}
\end{figure}

Next, we consider a step edge configuration  in which the layer 1 with the $d_{x^2-y^2}$ order parameter covers half of the long strip width as shown in Fig.~\ref{fig:0}(c). This type of edge is likely to occur in a real experimental setup where the twisted bilayer is fabricated using the cleave-and-stack technique.
Importantly, the edge of this type avoids the effects due to the pair breaking boundary by keeping the $d_{xy}$ order parameter nearly uniform. 

We model this configuration in the long strip geometry with a width of $N_y$ and maintain periodic boundary conditions along the $y$-direction in both layers. In layer 1, we impose a large potential barrier across half the strip, precluding the emergence of $d_{x^2-y^2}$ order parameter in the region with the barrier. We model this barrier as 
\begin{equation}\label{eq:barrier}
    V_{\textrm{barrier}}=
    \begin{cases}
      V_{\infty}, & \text{if}\ 0\leq y<0.25N_y \\
      0, & \text{if}\ 0.25N_y\leq y<0.75N_y \\
      V_{\infty}, & \text{if}\ 0.75N_y\leq y<N_y \\
    \end{cases}
\end{equation}
with $V_{\infty}\gg t$.

The self-consistent solution yields a $d_{xy}+id_{x^2-y^2}$ order parameter in the bulk. We obtain step edges of the $d_{x^2-y^2}$ order parameter in layer 1 at $y=0.25N_y$ and $y=0.75N_y$ while generating a nearly uniform $d_{xy}$ order parameter in layer 2 without any edges, as depicted in Fig.~\ref{fig:sc_step}. 

The pair-breaking edges of the $d_{xy}$ monolayer are absent in this configuration, leading to the reduction in the magnitude of the extended $s$-wave order parameter. 
The contribution of the $d_{xy}+is$ order parameter to the current in the system is negligible as we will demonstrate in Sec.~\ref{sec:currents}. Therefore, the step edge configuration can be used to examine the edge currents intrinsic to the chiral $d+id'$ superconductor formed in a twisted bilayer.

\subsubsection{Domain wall}
\begin{figure}
    \centering    
    \includegraphics[width=\linewidth]{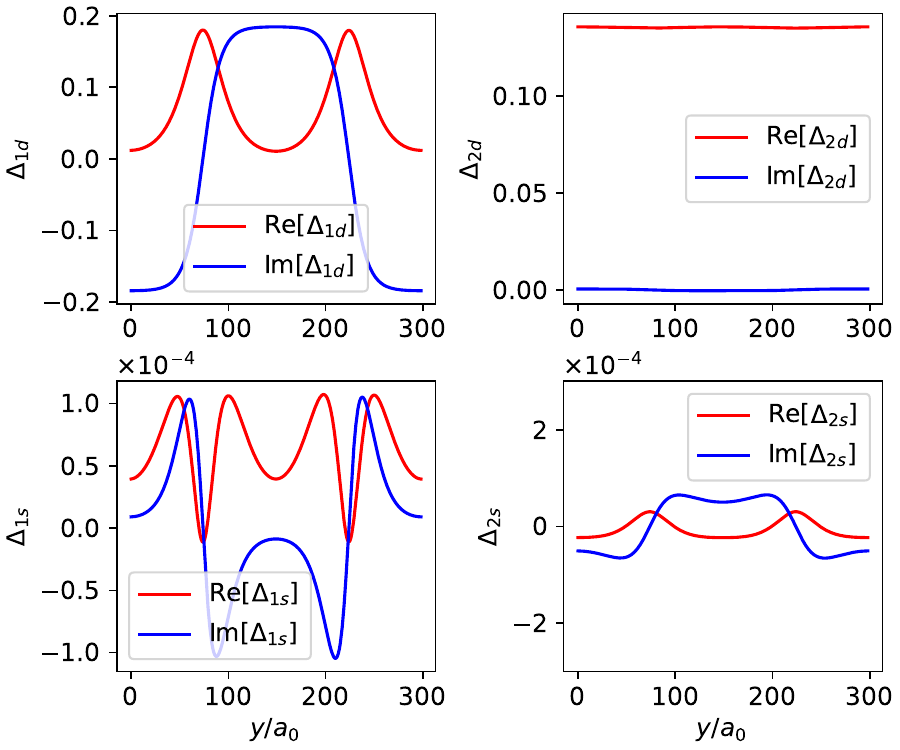}
    \caption[Spatial profile of the superconducting order parameters of the free domain wall configuration in the long strip geometry as a function of $y$.]{Spatial profile of the superconducting order parameters of the free domain wall configuration in the long strip geometry as a function of $y$. The top panel shows the self-consistent solutions of the $d-$wave order parameters in layers 1 and 2 labeled as $\Delta_{1d}$ ($d_{x^2-y^2}$) and $\Delta_{2d}$ ($d_{xy}$) respectively.
    The bottom panel shows the $s$-wave order parameters in layers 1 and 2, labeled as $\Delta_{1s}$ and $\Delta_{2s}$ respectively. The data depicts maximal superconducting gap at Fermi surface normalized with respect to $t$ for a weak inter-layer coupling ($g\approx8$~meV) and a temperature of $T/T_c\approx0.001$.}
    \label{fig:sc_domain1}
\end{figure}

Domain walls can form between $d+id'$ and $d-id'$ phases when there is a weak in-plane magnetic field modulating the relative phase or when the system is rapidly cooled. The Chern number for each domain is $\pm C$, respectively. A domain wall separating $d+id'$ and $d-id'$ regions therefore hosts twice as many edge modes compared to the ordinary edge, leading to a larger supercurrent. This configuration is illustrated in Fig.~\ref{fig:0}(d).

To avoid pair breaking effects we create domain walls in the $d_{x^2-y^2}$ order parameter in our system by imposing the following initial ansatz,
\begin{equation}
    \Delta_d=
    \begin{cases}
      d_{xy}-id_{x^2-y^2}, & \text{if}\ 0\leq y<0.25N_y, \\
      d_{xy}+id_{x^2-y^2}, & \text{if}\ 0.25N_y\leq y<0.75N_y, \\
      d_{xy}-id_{x^2-y^2}, & \text{if}\ 0.75N_y\leq y<N_y. \\
    \end{cases}
\end{equation}
Evidently, the domain walls are present at $y=0.25N_y$ and $y=0.75N_y$. We take periodic boundary conditions along the $y$-direction in both layers.  

In the bulk, our self-consistent solution Fig.~\ref{fig:sc_domain1} gives a nearly uniform $d_{xy}$ order parameter in layer 2 and, depending on the domain, a $\pm d_{x^2-y^2}$ order parameter in layer 1. In the absence of any pinning potential, we observe that the system prefers to transition smoothly from $d+id'$ to $d-id'$. The $d_{xy}$ order parameter remains nearly constant as there is no pair-breaking edge present in the system. The gradual change in the order parameter keeps the magnitude of the $d_{x^2-y^2}$ order parameter constant and vary the phase, thus creating a phase domain wall.

\begin{figure}
    \centering 
    \includegraphics[width=\linewidth]{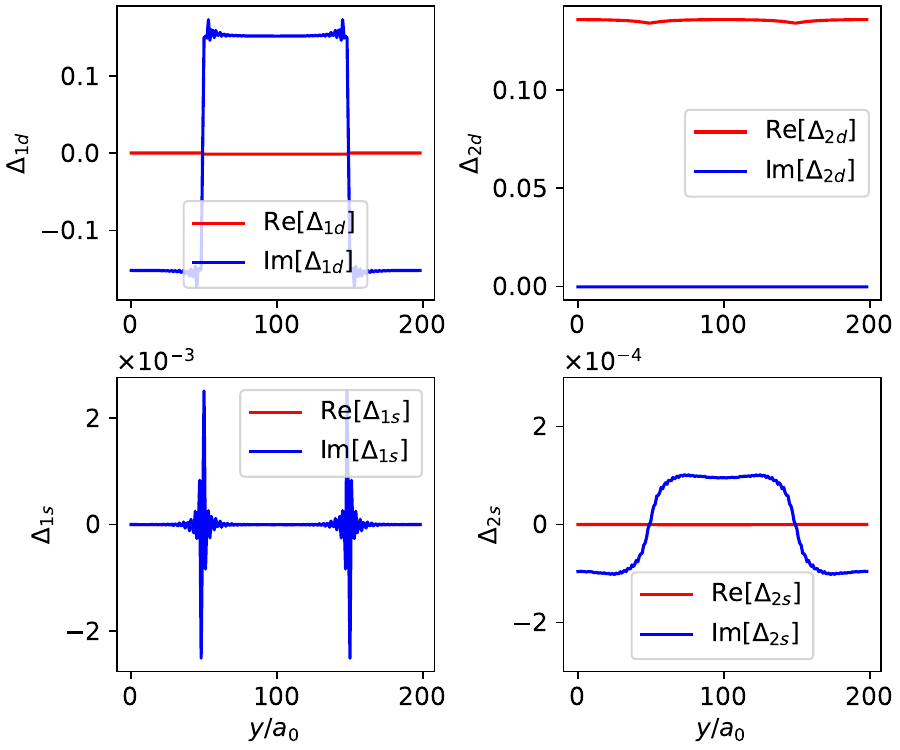}
    \caption[Spatial profile of the superconducting order parameters of the pinned domain wall configuration in the long strip geometry as a function of $y$.]{Spatial profile of the superconducting order parameters of the pinned domain wall configuration in the long strip geometry as a function of $y$. The top panel shows the self-consistent solutions of the $d-$wave order parameters in layers 1 and 2 labeled as $\Delta_{1d}$ ($d_{x^2-y^2}$) and $\Delta_{2d}$ ($d_{xy}$) respectively.
    The bottom panel shows the $s$-wave order parameters in layers 1 and 2, labeled as $\Delta_{1s}$ and $\Delta_{2s}$ respectively. The data depicts maximal superconducting gap at Fermi surface normalized with respect to $t$ for a weak inter-layer coupling ($g\approx8$~meV) and a temperature of $T/T_c\approx0.001$.}
    \label{fig:sc_domain2}
\end{figure}

We can also pin the domain walls using a potential to suppress the order parameters locally along a line. In this case, the magnitude of the $d_{x^2-y^2}$ order parameter goes to zero at the domain wall, creating an amplitude wall. We show the order parameters of the pinned domain wall in Fig.~\ref{fig:sc_domain2}.  

For the discussion that follows, we refer to the domain wall in the absence of pinning potential as `free domain wall' and that with a pinning potential as a `pinned domain wall'. The two domain walls exhibit notable differences in the spatial profile of the supercurrent as we will discuss in Sec.~\ref{sec:currents}.

\section{Edge currents}\label{sec:currents}
\subsection{Current operator}

The current flowing along a bond between sites $i$ and $j$ in the tight binding model Eq.\ \eqref{eq:h1} is most easily obtained by performing the Peierls substitution $t_{ij}\to t_{ij}\exp{(ieA_{ij}/\hbar)}$ where $A_{ij}$ is the vector potential integrated along the bond. The current operator then follows from 
\begin{equation}\label{curr1}
j_{ij,a}={\partial H\over\partial A_{ij}}\biggr|_{A=0}=-{et\over\hbar}\sum_{\sigma}\left(ic^\dagger_{i\sigma a}c_{j\sigma a}+{\rm h.c.}\right),    
\end{equation}
and the last expression holds when sites $i$ and $j$ are nearest neighbors.
Given the BdG eigenstates at temperature $T$ it is straightforward to evaluate the expected value $J_{ij,a}$ of the current operator Eq.\ \eqref{curr1} for any bond. By symmetry and current conservation, we only expect non-zero currents to flow along the in-plane bonds that are parallel to the edge.

\begin{figure*}
    \centering
    \includegraphics[width=0.99\linewidth]{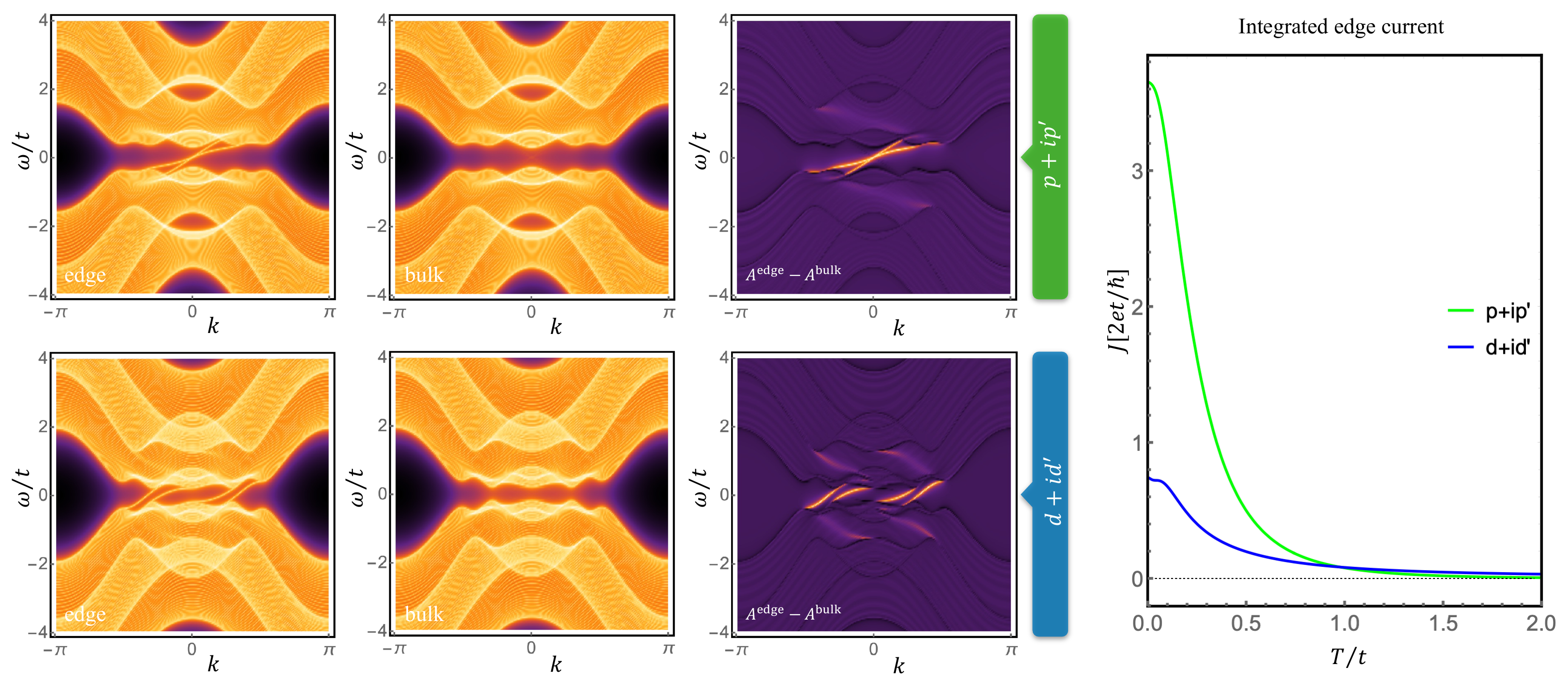}
    \caption[Spectral functions and the resulting integrated edge currents as a function of temperature $T$, for a $p+ip'$ and $d+id'$ superconductor on a long strip.]{Spectral functions Eq.\ \eqref{eq:spec} for a $p+ip'$ (top row) and $d+id'$ superconductor (bottom row) on a long strip of width $N_y=52$. The label ``edge" corresponds to a lattice site on the edge  whereas ``bulk" corresponds to a site near the strip center. In order to visualize various spectral features with maximum clarity we use larger gap $\Delta\simeq  0.5t$ and larger interlayer coupling $g\simeq 0.8t$. To model the $p+ip'$ SC we use Hamiltonian Eqs.\ (\ref{eq:h1},\ref{eq:h2}) adapted to describe a single flavor of spinless fermions and interactions that give rise to $p_x$ ($p_y$) order parameters in layers 1 (2). The rightmost panel shows integrated edge currents, as a function of temperature $T$ computed from Eq.\ \eqref{scurr}.}
    \label{fig:spec}
\end{figure*}

\subsection{Current from spectral function}

Before we present our results for edge current magnitudes we digress briefly to relate the edge current to the electron spectral function $A_k(y,\omega)$. The latter can be used to visualize the chiral edge modes and hence the relation offers a way to understand the origin of the edge current. We shall also see that this approach helps to rationalize the disparity between the robust edge currents expected in the $p+ip'$ ($m=1$) case and a relatively weak current predicted for the $d+id'$ ($m=2$) chiral superconductor. We remark that previous studies noted this disparity\ \cite{Huang2014, Tada2015, Suzuki2016, Nie2020}. Here, we provide a simple physical picture that would explain its origin.

For a long strip geometry where crystal momentum $k$ along $x$ is a good quantum number, starting from  Eq.\ \eqref{curr1} we can  derive a mixed representation expression for the current $J_{\hat{x}}(y)$ along a horizontal bond at distance $y$ from the edge in terms of the spectral function, 
\begin{equation}\label{scurr}
J_{\hat{x}}(y)=\frac{2et}{\hbar}\int_{-\pi}^\pi{dk\over 2\pi} \sin{k} \int_{-\infty}^\infty {d\omega\over 2\pi}\frac{\text{Tr}[A_k(y,\omega)]}{1+e^{\beta\omega}}.
\end{equation}
Here the trace extends over layer and BdG indices while 
\begin{equation}\label{eq:spec}
A_k(y,\omega)=-2{\rm Im}[\omega+i\delta-\cH(k)]_{yy}^{-1}
\end{equation}
is the spectral function evaluated at distance $y$ from the edge. $\cH(k)$ is the $4N_y\times 4N_y$ matrix Hamiltonian describing the strip and $\delta$ denotes a positive infinitesimal. Subscript $yy$ indicates that a diagonal $4\times 4$ block of the matrix at spatial position $y$ is to be taken. Individual elements of each such $4\times 4$ block can be thought of as position- and layer-resolved normal and anomalous components of the Gorkov Green's function $\hat{G}_k(y,y; \omega)$. Details of the derivation are given in Appendix A. 

Figure \ref{fig:spec} shows spectral fuctions $A_k(y,\omega)$ calculated from our bilayer model for both $p+ip'$ and $d+id'$ chiral superconductors. In accord with the expectations spectral functions show unidirectional modes when evaluated near the edge of the strip and they show fully gapped spectrum in the bulk. We note that for $p+ip'$ bilayer Chern number $C=1$ and 2 phases are possible, depending on the model parameters \cite{Tummuru2021}, while for $d+id'$ one can get $C=2,4$ \cite{Can2021}. We chose to display a $C=2$ phase for $p+ip'$ and $C=4$ for $d+id'$ here.

Because no current flows in the bulk of the strip one can subtract $A_k(y_B,\omega)$ from the edge spectral function in the argument of the trace in Eq.\ \eqref{scurr} without changing the result. Here $y_B$ denotes a position sufficiently far from the edge. Such bulk-subtracted edge spectral functions therefore help visualize the origin of the edge currents and are displayed in the third column of Fig.\ \ref{fig:spec}. We observe that for both $m=1$ and 2 the edge currents are sourced primarily from the edge modes, as expected. 

Furthermore, combined with Eq.\ \eqref{scurr} the spectral function plots help understand some important differences between $m=1$ and $m=2$ cases. First, we note that because of the Fermi function only the occupied states, that is the $\omega<0$ part of the spectral function, contribute to the current. The edge modes in the $p+ip'$ case therefore contribute only for $k<0$ and this contribution is weighted with the same sign by the $\sin{k}$ factor in Eq.\ \eqref{scurr}. Hence, after integration over $k$, we expect in this case a large contribution to the edge current. On the other hand the occupied edge modes in the $d+id'$ case have support at both positive and negative values of $k$; hence they will tend to cancel in the $k$-integral after being weighted by the odd $\sin{k}$ function. Indeed as illustrated in the rightmost panel of Fig.\ \ref{fig:spec} the integrated edge current calculated from Eq.\ \eqref{scurr} is about factor of 4 larger for $p+ip'$ than $d+id'$ for this specific choice of parameters. We note that, for the sake of simplicity, in this calculation the gap amplitudes are taken as $T$-independent and uniform across the width of the strip; the temperature dependence of $J$ comes entirely from the Fermi factor in Eq.\ \eqref{scurr}. Results for $J$ based on fully self-consistent calculations with realistic parameters will be presented in the next subsection.

\subsection{Edge currents: quantitative results}

Now, we calculate the edge currents in the long strip geometry of the $d+id'$ superconducting heterostructure based on fully self-consistent solutions of the BdG theory with parameters relevant to Bi-2212 bilayer. We compute the supercurrents along the periodic direction $x$ as a function of $y$ for the strip geometry with finite width along $y$ for the three current-carrying configurations described in Sec.~\ref{sec:configs}. We discuss the current density $I_{\hat{x}}(y)$ and use the net supercurrent associated with a given edge, that we define as $I_{\textrm{net}}=\sum_{y=0}^{N_y/2}I_{\hat{x}}(y)$, to quantify the edge currents.

\subsubsection{Aligned edge configuration}

\begin{figure}
    \centering   
    \includegraphics[width=\linewidth]{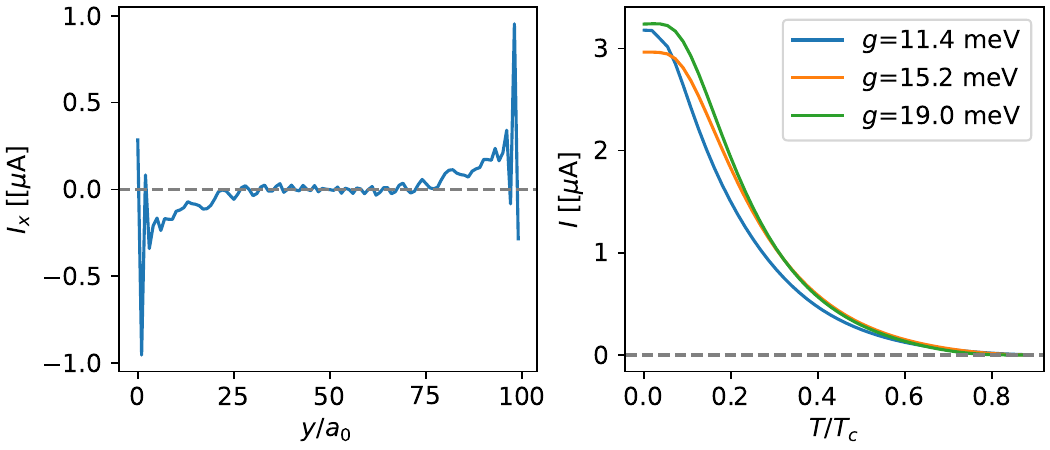}
    \caption[Edge currents in the aligned edge configuration.]{Edge currents in the aligned edge configuration. (Left) Spatial profile of the supercurrent flowing in $x$ as a function of $y$ for the aligned edge configuration at $g\approx11$~meV and $T/T_c\approx0.001$. (Right) Temperature dependence of the net edge currents for different inter-layer coupling values, $g$.}
    \label{fig:I_aligned}
\end{figure}

We begin with the simplest case, the aligned edge configuration, and present the supercurrent profile as a function of $y$ and the net supercurrent versus temperature in Fig.~\ref{fig:I_aligned}. For the aligned edge configuration, the supercurrents are localized in a narrow region at the common edges of both layers and decay exponentially into the bulk. For low temperatures, we observe that the magnitude of the net supercurrents in this configuration is significant for different values of interlayer couplings, $g$. However, upon decreasing the value of $g$, we do not observe any significant change in the magnitude of these supercurrents. We thus identify the origin of these supercurrents as the time-reversal breaking $d_{xy}+is$ order parameter nucleated at the pair-breaking edges. As discussed in the previous Section this effect is unrelated to the chiral $d+id'$ phase in twisted bilayer.

\subsubsection{Step edge configuration}

\begin{figure}
    \centering    
    \includegraphics[width=\linewidth]{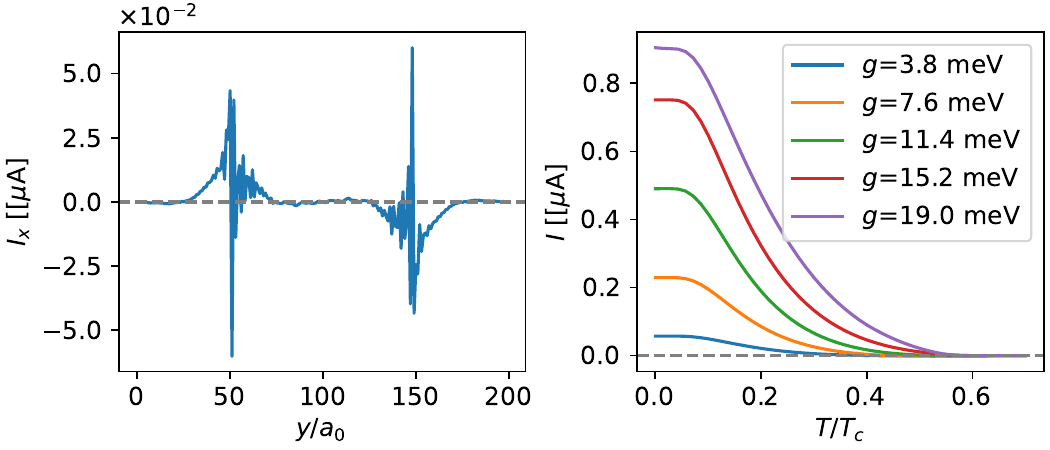}
    \caption[Edge currents in the step edge configuration.]{Edge currents in the step edge configuration. (Left) Spatial profile of the supercurrent flowing in $x$ as a function of $y$ for the step edge configuration at $g\approx8$~meV and $T/T_c\approx0.001$. (Right) Temperature dependence of the net edge currents for $g=4-19$~meV.}
    \label{fig:I_step}
\end{figure}

We now analyze the edge currents in the step edge configuration. Here, we avoid the pair-breaking edge of the $d_{xy}$ order parameter by considering step edges in the $d_{x^2-y^2}$ layer.  In Fig.~\ref{fig:I_step} we plot $I_{\hat{x}}(y)$ and analyze the net current as a function of temperature and interlayer coupling parameter.
 With the step edges of $d_{x^2-y^2}$ order parameter located at $y=0.25N_y$ and $y=0.75N_y$, we find that the supercurrents are maximal near these edges and decay into the bulk on either side of the edge. 

The net edge current is found to decrease with an increasing temperature. Further, as we decrease the inter-layer coupling $g$ from 19~meV to 4~meV, we find that the net supercurrent hosted in the system decreases at all temperatures. In the absence of interlayer coupling, i.e. at $g=0$, the supercurrents vanish. The supercurrents present in this system arise solely due to the time-reversal breaking of the $d+id'$ order parameter.

For temperatures below $0.1T_c$ and for $g=$4 to 19~meV, we estimate the net supercurrents hosted at the edge of the twisted cuprate bilayers to be between 0.10 to 0.9~$\mu$A for realistic parameters.

\subsubsection{Domain wall configuration}

\begin{figure}
    \centering 
    \includegraphics[width=\linewidth]{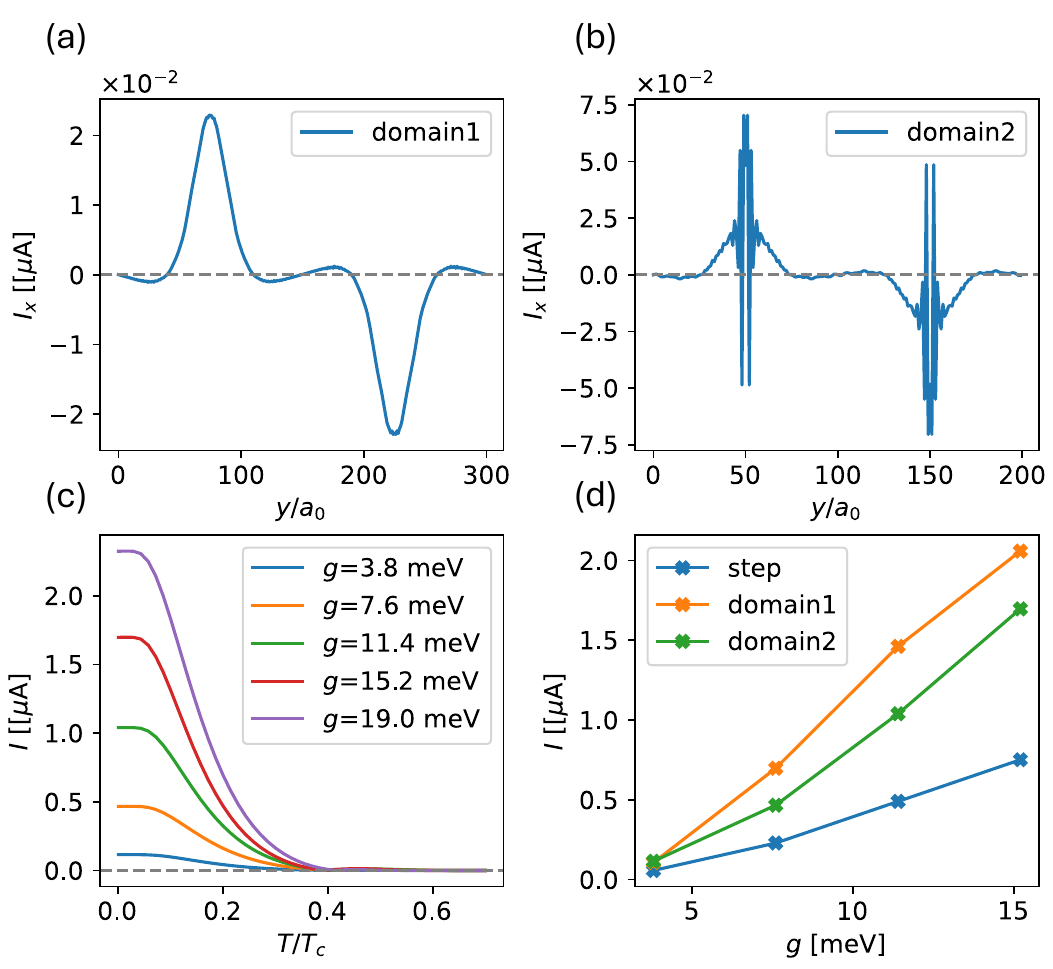}
    \caption[Edge currents in the domain wall configuration.]{Edge currents in the domain wall configuration. Spatial profile of the supercurrent flowing in $x$ as a function of $y$ for $g\approx8$~meV for: (a) the free domain wall (`domain1') configuration and (b) the pinned domain wall (`domain2'). (c) Temperature dependence of the net edge currents for the pinned domain wall configuration for different $g$ values. (d) Comparison of the net edge currents at the step edge, free domain wall (`domain1'), and the pinned domain wall (`domain2') as a function of $g$. $T/T_c<0.01$ for panels (a),(b),(d). }
    \label{fig:I_domain}
\end{figure}

Finally, we investigate supercurrents generated by the two types of a domain wall.  We show the current profile and the net current associated with the free domain wall as well as the pinned domain wall in Fig.~\ref{fig:I_domain}. 

In the pinned domain wall, the amplitude of the $d_{x^2-y^2}$ order parameter goes to zero, whereas it remains constant in the absence of pinning, where a phase domain wall is energetically more stable. As a result, the supercurrent profiles as a function of the position differ significantly for the two domain wall configurations. Supercurrents are highly localized and decay quickly into the bulk near the pinned domain wall, similar to the step edge configuration. Also similar to the step edge configuration, the supercurrents exhibit rapid oscillations akin to Friedel oscillations~\cite{tinkham}. In the absence of pinning, currents are still maximal at the domain wall but spread out significantly more into the bulk. These differences are related to the contrasting order parameter profiles for the two types of domain walls shown in Figs.\ \ref{fig:sc_domain1} and ~\ref{fig:sc_domain2}; the phase domain wall is clearly much broader than the amplitude domain wall.

Since the domain wall contains twice the number of protected chiral modes as compared to an edge, we expect the supercurrents to be larger than those generated by the step edge configuration. This is indeed the case, as the net current associated with both the domain walls is at least twice that of the step edge configuration. For the pinned domain wall, the peak values between $0.2-2.0$ $\mu$A for interlayer couplings ranging from $g=4$ meV to $19$ meV. The supercurrents reduce to zero as the interlayer coupling value decreases to zero, which is what we expect in the case of a $d+id'$ state in a bilayer.

Overall, the net supercurrents in pinned and free domain walls increase as the temperature decreases. As for the free domain wall, we observe that the domain wall thickness increases with temperature, leading to finite size effects in our model at higher temperatures. Nevertheless, our system sizes can accurately represent domain walls present in a $d+id'$ superconductor at low temperatures (below $0.01T_c$) relevant to experimental observations discussed below.

\section{Estimation of magnetic fields generated by edge currents}

The edge currents produced in a time-reversal breaking chiral superconductor will necessarily produce a magnetic field that can, in principle, be detected by state-of-the-art magnetic probes~\cite{Hartmann1999,Casola2018,Kirtley2010}. A possible way to detect the magnetic fields due to the edge currents is by using a scanning superconducting quantum interference device (SQUID) microscope. A DC SQUID consists of a superconducting loop, interrupted by two Josephson junctions.  The critical current and voltage drop across the device are periodic in the external magnetic flux.  The sensitivity and versatility of scanning SQUID have been instrumental in noninvasive measurements of a broad range of electronic orders such as those in unconventional superconductivity~\cite{Tsuei1994,Kalisky2011}, exotic magnetism~\cite{Christensen2019}, topological states~\cite{Uri2020}, and more~\cite{Persky2021}. 

As demonstrated in the previous Section currents generated by the step edge or pinned domain wall decay over tens of lattice sites, which corresponds to a lengthscale of several nm. A typical magnetic field sensor is located at a height $h>20$ nm above the sample. In this geometry, the edge current can be treated as a line current concentrated at the edge for all practical purposes~\cite{Kirtley2010}. We thus model the edge as a thin long wire along $x$ carrying the net current $I_{\rm net}$, located at $(y,z)=(0,0)$. The magnetic field generated by this line current at height $h$ above the sample can be deduced from Amp\'ere's law and is given by
\begin{align}\label{eq:Br}
    \mathbf{B}&=\frac{\mu_0 I_{\rm net}}{2\pi r}\hat{\theta} =\frac{\mu_0 I_{\rm net}}{2\pi r} \left(\frac{y}{r} \hat{z} - \frac{h}{r} \hat{y}\right),
\end{align}
where $y$ denotes the lateral position of the sensor and $r=\sqrt{y^2+h^2}$ is the radial distance from the wire. 

The flux picked up by the SQUID loop arises from the $z-$component of the magnetic field in Eq.~\eqref{eq:Br}. The minimal sensor-sample distance depends on the SQUID design which typically balances spatial resolution and magnetic field sensitivity~\cite{Persky2021,Kirtley2010}.  For planar SQUID, this distance is 100-500 nm~\cite{Kirtley2016,Cui2017}, while for SQUID-on-tip and nano-SQUID, it is 20-100 nm~\cite{Vasyukov2013,Uri2020}.

\begin{figure}
    \centering
    \includegraphics[width=\linewidth]{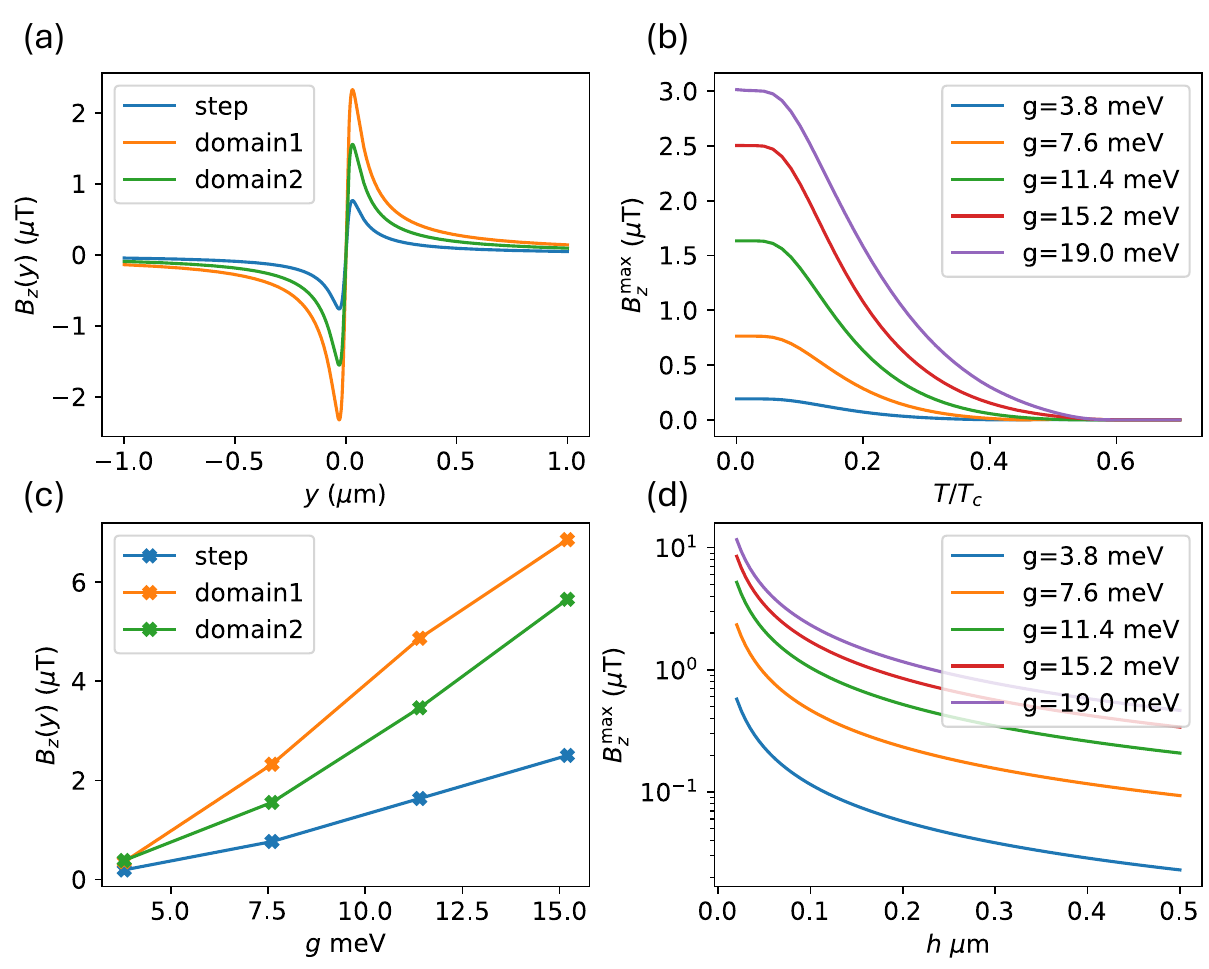}
    \caption[Estimation of magnetic fields due to edge currents.]{Estimation of magnetic fields due to edge currents. (a) Perpendicular magnetic field ($B_z$) generated due to edge currents as a function of $y$ at a vertical height of $h=30$~nm for step edge and domain walls. Here, the net edge currents are modeled as a line current flowing along $x$ located at $(y,z)=(0,0)$. (b) Maximum perpendicular magnetic field ($B_z^{\textrm{max}}$) due to the net currents in the step edge configuration as a function of temperature at $h=30$~nm. (c) Comparison of $B_z^{\textrm{max}}$ ($T/T_c<0.01$) at the step edge and domain walls as a function of $g$ at $h=30$~nm. (d) $B_z^{\textrm{max}}$ for the pinned domain wall configuration as a function of height.}
    \label{fig:magnetic_fields}
\end{figure}

We calculate the magnetic fields generated from the step edge and the domain wall edge of the $d+id'$ bilayer. In Fig.~\ref{fig:magnetic_fields}, we present the magnetic field profiles for the step edge and the pinned domain wall at a height of 30 nm from the sample, as a function of $y$.  We also plot the peak magnetic field generated at the step edge as a function of temperature and confirm that the peak magnetic field shows similar features as the supercurrent plot in Fig.~\ref{fig:I_step}. Additionally, we plot the peak magnetic field for the step edge and domain wall edges as a function of interlayer coupling. 

Our estimates indicate that a magnetic sensor positioned at a height of 30 nm from the sample would detect magnetic fields of 0.1-2 $\mu$T at the step edge, 0.2-5 $\mu$T at the domain walls, for realistic interlayer coupling values below 15~meV. These values are well above the detection threshold of current state-of-the-art SQUIDS such as nano-SQUIDs~\cite{Uri2020}.

We also show in Fig.~\ref{fig:magnetic_fields} the variation of the peak magnetic field with the vertical height of the sample for a representative edge (pinned domain wall). The order of magnitude of the magnetic field remains above $1$ $\mu$T below a height of 100 nm which is well above the detection threshold of nano-SQUIDs. The magnetic field decreases to $\sim 100$ nT for realistic values of interlayer coupling going from 100 to 500 nm height where planar SQUIDs typically operate.

Currents associated with the phase domain wall are spread out over a wider lengthscale so the the above thin-wire approximation becomes less accurate. Nevertheless, this prescription should still provide a reasonable estimate of the magnetic field strength, which is expected to be again between 1-5 $\mu$T. 


\section{Discussion and summary}
\label{sec:conclusion}
The twisted bilayer cuprate system breaks time-reversal symmetry due to the formation of $d+id'$ superconducting order parameter. Our work analyzes the edge currents that arise in such a  chiral superconductor and can be probed experimentally as a signature of the chiral phase. 

We study this system using an aligned lattice model where the $d$-wave superconducting pairing occurs on the nearest-neighbor bonds in one layer and on the next-nearest neighbor bonds in the second layer, mimicking the $d_{x^2-y^2}$ and the $d_{xy}$ order parameters respectively while avoiding the complications arising from large moir\'e unit cell. Working with the long strip geometry periodic along the $x$-direction and a finite width along the $y$-direction we study three different edge configurations and calculate the supercurrents flowing in each of these geometries. We also relate the edge currents to the spectral function in a generic chiral superconductor and provide an intuitive picture for their difference in magnitudes in the $p+ip'$  and a $d+id'$ case. We then use realistic parameters to model the supercurrents at the edge of the twisted cuprate bilayers using a fully self-consistent Bogoliubov-de Gennes theory.

The aligned edge configuration corresponds to the scenario where the edges of the two layers are perfectly aligned. In accord with previous work \cite{Hakansson2015,Holmvall2018,Holmvall2019,Wennerdal2020} we find that a significant extended $s$-wave order parameter develops at the pair-breaking edge of the $d_{xy}$ monolayer resulting in the nucleation of a time-reversal breaking $d_{xy}+is$ order parameter locally at the edge. We further find that the main contribution to the current for this edge configuration arises from this $d+is$ order parameter, which is in itself interesting but unrelated to the bulk $d+id'$ phase of interest here.  To study the edge currents and resultant magnetic fields arising solely due to the bulk $d+id'$ order we therefore focus on two other types of linear defects, the step edge and the domain wall, chosen such that no significant $s$ component is nucleated. 

The step edge configuration features a sharp boundary in the $d_{x^2-y^2}$ order parameter but avoids the pair-breaking edge in the $d_{xy}$ monolayer.  The supercurrents at these step edges arise solely due to the bulk $d+id'$ order parameter. For realistic parameters relevant to twisted bilayer cuprates, we estimate the supercurrents from this configuration to be 0.10-0.9 $\mu$A. Near domain walls we observe supercurrents that are at least twice the magnitude of those present at the step edges, with some notable differences in the local current density profiles between free and pinned domain walls.

Based on these results we provide estimates of the magnetic fields generated by the supercurrents at the edges. At the height of 30 nm above the sample, characteristic of the nanoSQUID experiment, we estimate the peak magnetic fields to be of the order of 1 $\mu$T generated by a step edge or pinned domain wall configuration in the $d+id'$ superconducting phase. This places the edge currents in twisted Bi$_2$Sr$_2$CaCu$_2$O$_{8+x}$ bilayers within the realm of what is currently observable using the state of the art scanning magnetometry. The principal source of uncertainty entering the field strength estimate is the magnitude of the interlayer tunneling amplitude $g$ on which the edge current depends strongly. The range of values quoted above corresponds to the range of $g=4-15$ meV spanning its most likely value consistent with transport measurements \cite{Tummuru2022,Zhao2021}. By contrast, the edge current shows only weak dependence on other system features, such as the detailed shape of the Fermi surface and the size of the gap. The parameters underlying these features are also much better known for Bi$_2$Sr$_2$CaCu$_2$O$_{8+x}$ and other cuprates.

In closing we wish to remark on the general significance of these findings. As of this writing twisted cuprate bilayers constitute the most plausible candidate for a topological superconductor in two dimensions and the sole known candidate for a high-temperature topological superconductor. Ref.\ \cite{Zhao2021} provided strong evidence for spontaneous $\cT$ breaking in high-quality Bi-2212 junctions near $45^\circ$ twist which is a basic pre-requisite for the chiral $d+id'$ phase. However, to directly probe topology it is necessary to detect the protected chiral edge modes which is a more delicate task. Edge modes in a topological superconductor carry quantized heat current but this is exceedingly difficult to measure. The associated persistent electrical current is not quantized but may be more easily observed because it generates magnetic field. We find its magnitude to be in the range of a few $\mu T$, which is above the detection threshold of modern magnetometry techniques. As we emphasized, the amplitude of the current is non-universal and exhibits fairly significant dependence on the edge configuration; the above estimates pertain to the simplest types of edges that we expect to occur in physical systems.   
An important caveat will be to stay away from pair-breaking edges which tend to produce even larger edge currents through a physically distinct mechanism that applies already in the monolayer limit. Such currents are independently interesting but are unrelated to the chiral $d+id'$ phase that is only expected to occur in twisted bilayers. 

Detection of the edge currents in twisted cuprate bilayers thus poses a significant experimental challenge, but it is a challenge well worth pursuing. The high-temperature topological phase predicted to occur near the $45^\circ$ twist has been extensively discussed in the literature as basis for a number of novel phenomena and applications. These include charge-4$e$ superconductivity \cite{Liu2023}, higher order topology \cite{Li2023}, fractional and coreless vortices \cite{Holmvall2023} and Majorana fermions \cite{Margalit2022,Mercado2022}. In addition, the idea of assembling superconducting monolayers with a twist has numerous interesting applications outside the realm of high-$T_c$ cuprates; these include twisted iron-based bilayers \cite{Vafek2023}, non-abelian topology in twisted spin-singlet valley-triplet superconductors \cite{Benjamin2023} and related innovative proposals \cite{Fan2023,Lin_2023}, as well as a proposed application to improve the performance of transmon qubits \cite{Brosco2024,Patel2024} which power the large majority of superconducting quantum processors currently in operation.   

\section{Acknowledgments}
We thank Catherine Kallin, Thomas Scaffidi, Alberto Nocera, Niclas Heinsdorff, Nitin Kaushal and Aviram Uri for helpful discussions and correspondence. This work was supported by NSERC, CIFAR and the Canada First Research Excellence Fund, Quantum Materials and Future Technologies Program. M.F. thanks the Aspen Center for Physics where part of this work was completed.

\bibliography{edgecurrents}

\begin{thebibliography}{57}%
\makeatletter
\providecommand \@ifxundefined [1]{%
 \@ifx{#1\undefined}
}%
\providecommand \@ifnum [1]{%
 \ifnum #1\expandafter \@firstoftwo
 \else \expandafter \@secondoftwo
 \fi
}%
\providecommand \@ifx [1]{%
 \ifx #1\expandafter \@firstoftwo
 \else \expandafter \@secondoftwo
 \fi
}%
\providecommand \natexlab [1]{#1}%
\providecommand \enquote  [1]{``#1''}%
\providecommand \bibnamefont  [1]{#1}%
\providecommand \bibfnamefont [1]{#1}%
\providecommand \citenamefont [1]{#1}%
\providecommand \href@noop [0]{\@secondoftwo}%
\providecommand \href [0]{\begingroup \@sanitize@url \@href}%
\providecommand \@href[1]{\@@startlink{#1}\@@href}%
\providecommand \@@href[1]{\endgroup#1\@@endlink}%
\providecommand \@sanitize@url [0]{\catcode `\\12\catcode `\$12\catcode `\&12\catcode `\#12\catcode `\^12\catcode `\_12\catcode `\%12\relax}%
\providecommand \@@startlink[1]{}%
\providecommand \@@endlink[0]{}%
\providecommand \url  [0]{\begingroup\@sanitize@url \@url }%
\providecommand \@url [1]{\endgroup\@href {#1}{\urlprefix }}%
\providecommand \urlprefix  [0]{URL }%
\providecommand \Eprint [0]{\href }%
\providecommand \doibase [0]{https://doi.org/}%
\providecommand \selectlanguage [0]{\@gobble}%
\providecommand \bibinfo  [0]{\@secondoftwo}%
\providecommand \bibfield  [0]{\@secondoftwo}%
\providecommand \translation [1]{[#1]}%
\providecommand \BibitemOpen [0]{}%
\providecommand \bibitemStop [0]{}%
\providecommand \bibitemNoStop [0]{.\EOS\space}%
\providecommand \EOS [0]{\spacefactor3000\relax}%
\providecommand \BibitemShut  [1]{\csname bibitem#1\endcsname}%
\let\auto@bib@innerbib\@empty
\bibitem [{\citenamefont {Stone}\ and\ \citenamefont {Roy}(2004)}]{Stone2004}%
  \BibitemOpen
  \bibfield  {author} {\bibinfo {author} {\bibfnamefont {M.}~\bibnamefont {Stone}}\ and\ \bibinfo {author} {\bibfnamefont {R.}~\bibnamefont {Roy}},\ }\bibfield  {title} {\bibinfo {title} {Edge modes, edge currents, and gauge invariance in ${p}_{x}{+ip}_{y}$ superfluids and superconductors},\ }\href {https://doi.org/10.1103/PhysRevB.69.184511} {\bibfield  {journal} {\bibinfo  {journal} {Phys. Rev. B}\ }\textbf {\bibinfo {volume} {69}},\ \bibinfo {pages} {184511} (\bibinfo {year} {2004})}\BibitemShut {NoStop}%
\bibitem [{\citenamefont {Huang}\ \emph {et~al.}(2014)\citenamefont {Huang}, \citenamefont {Taylor},\ and\ \citenamefont {Kallin}}]{Huang2014}%
  \BibitemOpen
  \bibfield  {author} {\bibinfo {author} {\bibfnamefont {W.}~\bibnamefont {Huang}}, \bibinfo {author} {\bibfnamefont {E.}~\bibnamefont {Taylor}},\ and\ \bibinfo {author} {\bibfnamefont {C.}~\bibnamefont {Kallin}},\ }\bibfield  {title} {\bibinfo {title} {Vanishing edge currents in non-$p$-wave topological chiral superconductors},\ }\href {https://doi.org/10.1103/PhysRevB.90.224519} {\bibfield  {journal} {\bibinfo  {journal} {Phys. Rev. B}\ }\textbf {\bibinfo {volume} {90}},\ \bibinfo {pages} {224519} (\bibinfo {year} {2014})}\BibitemShut {NoStop}%
\bibitem [{\citenamefont {Kallin}\ and\ \citenamefont {Berlinsky}(2016)}]{Kallin2016}%
  \BibitemOpen
  \bibfield  {author} {\bibinfo {author} {\bibfnamefont {C.}~\bibnamefont {Kallin}}\ and\ \bibinfo {author} {\bibfnamefont {J.}~\bibnamefont {Berlinsky}},\ }\bibfield  {title} {\bibinfo {title} {Chiral superconductors},\ }\href {https://doi.org/10.1088/0034-4885/79/5/054502} {\bibfield  {journal} {\bibinfo  {journal} {Reports on Progress in Physics}\ }\textbf {\bibinfo {volume} {79}},\ \bibinfo {pages} {054502} (\bibinfo {year} {2016})}\BibitemShut {NoStop}%
\bibitem [{\citenamefont {Huang}\ \emph {et~al.}(2015)\citenamefont {Huang}, \citenamefont {Lederer}, \citenamefont {Taylor},\ and\ \citenamefont {Kallin}}]{Huang2015}%
  \BibitemOpen
  \bibfield  {author} {\bibinfo {author} {\bibfnamefont {W.}~\bibnamefont {Huang}}, \bibinfo {author} {\bibfnamefont {S.}~\bibnamefont {Lederer}}, \bibinfo {author} {\bibfnamefont {E.}~\bibnamefont {Taylor}},\ and\ \bibinfo {author} {\bibfnamefont {C.}~\bibnamefont {Kallin}},\ }\bibfield  {title} {\bibinfo {title} {Nontopological nature of the edge current in a chiral $p$-wave superconductor},\ }\href {https://doi.org/10.1103/PhysRevB.91.094507} {\bibfield  {journal} {\bibinfo  {journal} {Phys. Rev. B}\ }\textbf {\bibinfo {volume} {91}},\ \bibinfo {pages} {094507} (\bibinfo {year} {2015})}\BibitemShut {NoStop}%
\bibitem [{\citenamefont {Tada}\ \emph {et~al.}(2015)\citenamefont {Tada}, \citenamefont {Nie},\ and\ \citenamefont {Oshikawa}}]{Tada2015}%
  \BibitemOpen
  \bibfield  {author} {\bibinfo {author} {\bibfnamefont {Y.}~\bibnamefont {Tada}}, \bibinfo {author} {\bibfnamefont {W.}~\bibnamefont {Nie}},\ and\ \bibinfo {author} {\bibfnamefont {M.}~\bibnamefont {Oshikawa}},\ }\bibfield  {title} {\bibinfo {title} {Orbital angular momentum and spectral flow in two-dimensional chiral superfluids},\ }\href {https://doi.org/10.1103/PhysRevLett.114.195301} {\bibfield  {journal} {\bibinfo  {journal} {Phys. Rev. Lett.}\ }\textbf {\bibinfo {volume} {114}},\ \bibinfo {pages} {195301} (\bibinfo {year} {2015})}\BibitemShut {NoStop}%
\bibitem [{\citenamefont {Wang}\ \emph {et~al.}(2018)\citenamefont {Wang}, \citenamefont {Wang},\ and\ \citenamefont {Kallin}}]{Wang2018}%
  \BibitemOpen
  \bibfield  {author} {\bibinfo {author} {\bibfnamefont {X.}~\bibnamefont {Wang}}, \bibinfo {author} {\bibfnamefont {Z.}~\bibnamefont {Wang}},\ and\ \bibinfo {author} {\bibfnamefont {C.}~\bibnamefont {Kallin}},\ }\bibfield  {title} {\bibinfo {title} {Spontaneous edge current in higher chirality superconductors},\ }\href {https://doi.org/10.1103/PhysRevB.98.094501} {\bibfield  {journal} {\bibinfo  {journal} {Phys. Rev. B}\ }\textbf {\bibinfo {volume} {98}},\ \bibinfo {pages} {094501} (\bibinfo {year} {2018})}\BibitemShut {NoStop}%
\bibitem [{\citenamefont {Rainer}\ \emph {et~al.}(1998)\citenamefont {Rainer}, \citenamefont {Burkhardt}, \citenamefont {Fogelström},\ and\ \citenamefont {Sauls}}]{Rainer1998}%
  \BibitemOpen
  \bibfield  {author} {\bibinfo {author} {\bibfnamefont {D.}~\bibnamefont {Rainer}}, \bibinfo {author} {\bibfnamefont {H.}~\bibnamefont {Burkhardt}}, \bibinfo {author} {\bibfnamefont {M.}~\bibnamefont {Fogelström}},\ and\ \bibinfo {author} {\bibfnamefont {J.}~\bibnamefont {Sauls}},\ }\bibfield  {title} {\bibinfo {title} {Andreev bound states, surfaces and subdominant pairing in high tc superconductors},\ }\href {https://doi.org/https://doi.org/10.1016/S0022-3697(98)00175-9} {\bibfield  {journal} {\bibinfo  {journal} {Journal of Physics and Chemistry of Solids}\ }\textbf {\bibinfo {volume} {59}},\ \bibinfo {pages} {2040} (\bibinfo {year} {1998})}\BibitemShut {NoStop}%
\bibitem [{\citenamefont {Horovitz}\ and\ \citenamefont {Golub}(2003)}]{Horovitz2003}%
  \BibitemOpen
  \bibfield  {author} {\bibinfo {author} {\bibfnamefont {B.}~\bibnamefont {Horovitz}}\ and\ \bibinfo {author} {\bibfnamefont {A.}~\bibnamefont {Golub}},\ }\bibfield  {title} {\bibinfo {title} {Superconductors with broken time-reversal symmetry: Spontaneous magnetization and quantum hall effects},\ }\href {https://doi.org/10.1103/PhysRevB.68.214503} {\bibfield  {journal} {\bibinfo  {journal} {Phys. Rev. B}\ }\textbf {\bibinfo {volume} {68}},\ \bibinfo {pages} {214503} (\bibinfo {year} {2003})}\BibitemShut {NoStop}%
\bibitem [{\citenamefont {Braunecker}\ \emph {et~al.}(2005)\citenamefont {Braunecker}, \citenamefont {Lee},\ and\ \citenamefont {Wang}}]{Braunecker2005}%
  \BibitemOpen
  \bibfield  {author} {\bibinfo {author} {\bibfnamefont {B.}~\bibnamefont {Braunecker}}, \bibinfo {author} {\bibfnamefont {P.~A.}\ \bibnamefont {Lee}},\ and\ \bibinfo {author} {\bibfnamefont {Z.}~\bibnamefont {Wang}},\ }\bibfield  {title} {\bibinfo {title} {Edge currents in superconductors with a broken time-reversal symmetry},\ }\href {https://doi.org/10.1103/PhysRevLett.95.017004} {\bibfield  {journal} {\bibinfo  {journal} {Phys. Rev. Lett.}\ }\textbf {\bibinfo {volume} {95}},\ \bibinfo {pages} {017004} (\bibinfo {year} {2005})}\BibitemShut {NoStop}%
\bibitem [{\citenamefont {Black-Schaffer}(2012)}]{Black-Schaffer2012}%
  \BibitemOpen
  \bibfield  {author} {\bibinfo {author} {\bibfnamefont {A.~M.}\ \bibnamefont {Black-Schaffer}},\ }\bibfield  {title} {\bibinfo {title} {Edge properties and majorana fermions in the proposed chiral $d$-wave superconducting state of doped graphene},\ }\href {https://doi.org/10.1103/PhysRevLett.109.197001} {\bibfield  {journal} {\bibinfo  {journal} {Phys. Rev. Lett.}\ }\textbf {\bibinfo {volume} {109}},\ \bibinfo {pages} {197001} (\bibinfo {year} {2012})}\BibitemShut {NoStop}%
\bibitem [{\citenamefont {Suzuki}\ and\ \citenamefont {Asano}(2016)}]{Suzuki2016}%
  \BibitemOpen
  \bibfield  {author} {\bibinfo {author} {\bibfnamefont {S.-I.}\ \bibnamefont {Suzuki}}\ and\ \bibinfo {author} {\bibfnamefont {Y.}~\bibnamefont {Asano}},\ }\bibfield  {title} {\bibinfo {title} {Spontaneous edge current in a small chiral superconductor with a rough surface},\ }\href {https://doi.org/10.1103/PhysRevB.94.155302} {\bibfield  {journal} {\bibinfo  {journal} {Phys. Rev. B}\ }\textbf {\bibinfo {volume} {94}},\ \bibinfo {pages} {155302} (\bibinfo {year} {2016})}\BibitemShut {NoStop}%
\bibitem [{\citenamefont {Holmvall}\ and\ \citenamefont {Black-Schaffer}(2023)}]{Holmvall2023_2}%
  \BibitemOpen
  \bibfield  {author} {\bibinfo {author} {\bibfnamefont {P.}~\bibnamefont {Holmvall}}\ and\ \bibinfo {author} {\bibfnamefont {A.~M.}\ \bibnamefont {Black-Schaffer}},\ }\bibfield  {title} {\bibinfo {title} {Enhanced chiral edge currents and orbital magnetic moment in chiral $d$-wave superconductors from mesoscopic finite-size effects},\ }\href {https://doi.org/10.1103/PhysRevB.108.174505} {\bibfield  {journal} {\bibinfo  {journal} {Phys. Rev. B}\ }\textbf {\bibinfo {volume} {108}},\ \bibinfo {pages} {174505} (\bibinfo {year} {2023})}\BibitemShut {NoStop}%
\bibitem [{\citenamefont {Can}\ \emph {et~al.}(2021{\natexlab{a}})\citenamefont {Can}, \citenamefont {Tummuru}, \citenamefont {Day}, \citenamefont {Elfimov}, \citenamefont {Damascelli},\ and\ \citenamefont {Franz}}]{Can2021}%
  \BibitemOpen
  \bibfield  {author} {\bibinfo {author} {\bibfnamefont {O.}~\bibnamefont {Can}}, \bibinfo {author} {\bibfnamefont {T.}~\bibnamefont {Tummuru}}, \bibinfo {author} {\bibfnamefont {R.~P.}\ \bibnamefont {Day}}, \bibinfo {author} {\bibfnamefont {I.}~\bibnamefont {Elfimov}}, \bibinfo {author} {\bibfnamefont {A.}~\bibnamefont {Damascelli}},\ and\ \bibinfo {author} {\bibfnamefont {M.}~\bibnamefont {Franz}},\ }\bibfield  {title} {\bibinfo {title} {High-temperature topological superconductivity in twisted double-layer copper oxides},\ }\href {https://doi.org/10.1038/s41567-020-01142-7} {\bibfield  {journal} {\bibinfo  {journal} {Nature Physics}\ }\textbf {\bibinfo {volume} {17}},\ \bibinfo {pages} {519} (\bibinfo {year} {2021}{\natexlab{a}})}\BibitemShut {NoStop}%
\bibitem [{\citenamefont {Zhao}\ \emph {et~al.}(2023)\citenamefont {Zhao}, \citenamefont {Cui}, \citenamefont {Volkov}, \citenamefont {Yoo}, \citenamefont {Lee}, \citenamefont {Gardener}, \citenamefont {Akey}, \citenamefont {Engelke}, \citenamefont {Ronen}, \citenamefont {Zhong}, \citenamefont {Gu}, \citenamefont {Plugge}, \citenamefont {Tummuru}, \citenamefont {Kim}, \citenamefont {Franz}, \citenamefont {Pixley}, \citenamefont {Poccia},\ and\ \citenamefont {Kim}}]{Zhao2021}%
  \BibitemOpen
  \bibfield  {author} {\bibinfo {author} {\bibfnamefont {S.~Y.~F.}\ \bibnamefont {Zhao}}, \bibinfo {author} {\bibfnamefont {X.}~\bibnamefont {Cui}}, \bibinfo {author} {\bibfnamefont {P.~A.}\ \bibnamefont {Volkov}}, \bibinfo {author} {\bibfnamefont {H.}~\bibnamefont {Yoo}}, \bibinfo {author} {\bibfnamefont {S.}~\bibnamefont {Lee}}, \bibinfo {author} {\bibfnamefont {J.~A.}\ \bibnamefont {Gardener}}, \bibinfo {author} {\bibfnamefont {A.~J.}\ \bibnamefont {Akey}}, \bibinfo {author} {\bibfnamefont {R.}~\bibnamefont {Engelke}}, \bibinfo {author} {\bibfnamefont {Y.}~\bibnamefont {Ronen}}, \bibinfo {author} {\bibfnamefont {R.}~\bibnamefont {Zhong}}, \bibinfo {author} {\bibfnamefont {G.}~\bibnamefont {Gu}}, \bibinfo {author} {\bibfnamefont {S.}~\bibnamefont {Plugge}}, \bibinfo {author} {\bibfnamefont {T.}~\bibnamefont {Tummuru}}, \bibinfo {author} {\bibfnamefont {M.}~\bibnamefont {Kim}}, \bibinfo {author} {\bibfnamefont {M.}~\bibnamefont {Franz}}, \bibinfo {author} {\bibfnamefont {J.~H.}\ \bibnamefont {Pixley}},
  \bibinfo {author} {\bibfnamefont {N.}~\bibnamefont {Poccia}},\ and\ \bibinfo {author} {\bibfnamefont {P.}~\bibnamefont {Kim}},\ }\bibfield  {title} {\bibinfo {title} {Time-reversal symmetry breaking superconductivity between twisted cuprate superconductors},\ }\href {https://doi.org/10.1126/science.abl8371} {\bibfield  {journal} {\bibinfo  {journal} {Science}\ }\textbf {\bibinfo {volume} {382}},\ \bibinfo {pages} {1422} (\bibinfo {year} {2023})}\BibitemShut {NoStop}%
\bibitem [{\citenamefont {Volkov}\ \emph {et~al.}(2023{\natexlab{a}})\citenamefont {Volkov}, \citenamefont {Étienne Lantagne-Hurtubise}, \citenamefont {Tummuru}, \citenamefont {Plugge}, \citenamefont {Pixley},\ and\ \citenamefont {Franz}}]{Etienne2023}%
  \BibitemOpen
  \bibfield  {author} {\bibinfo {author} {\bibfnamefont {P.~A.}\ \bibnamefont {Volkov}}, \bibinfo {author} {\bibnamefont {Étienne Lantagne-Hurtubise}}, \bibinfo {author} {\bibfnamefont {T.}~\bibnamefont {Tummuru}}, \bibinfo {author} {\bibfnamefont {S.}~\bibnamefont {Plugge}}, \bibinfo {author} {\bibfnamefont {J.~H.}\ \bibnamefont {Pixley}},\ and\ \bibinfo {author} {\bibfnamefont {M.}~\bibnamefont {Franz}},\ }\href@noop {} {\bibinfo {title} {Josephson diode effects in twisted nodal superconductors}} (\bibinfo {year} {2023}{\natexlab{a}}),\ \Eprint {https://arxiv.org/abs/2307.01261} {arXiv:2307.01261 [cond-mat.supr-con]} \BibitemShut {NoStop}%
\bibitem [{\citenamefont {Zhu}\ \emph {et~al.}(2021)\citenamefont {Zhu}, \citenamefont {Liao}, \citenamefont {Zhang}, \citenamefont {Xie}, \citenamefont {Meng}, \citenamefont {Liu}, \citenamefont {Bai}, \citenamefont {Ji}, \citenamefont {Zhang}, \citenamefont {Jiang}, \citenamefont {Zhong}, \citenamefont {Schneeloch}, \citenamefont {Gu}, \citenamefont {Gu}, \citenamefont {Ma}, \citenamefont {Zhang},\ and\ \citenamefont {Xue}}]{Xue2021}%
  \BibitemOpen
  \bibfield  {author} {\bibinfo {author} {\bibfnamefont {Y.}~\bibnamefont {Zhu}}, \bibinfo {author} {\bibfnamefont {M.}~\bibnamefont {Liao}}, \bibinfo {author} {\bibfnamefont {Q.}~\bibnamefont {Zhang}}, \bibinfo {author} {\bibfnamefont {H.-Y.}\ \bibnamefont {Xie}}, \bibinfo {author} {\bibfnamefont {F.}~\bibnamefont {Meng}}, \bibinfo {author} {\bibfnamefont {Y.}~\bibnamefont {Liu}}, \bibinfo {author} {\bibfnamefont {Z.}~\bibnamefont {Bai}}, \bibinfo {author} {\bibfnamefont {S.}~\bibnamefont {Ji}}, \bibinfo {author} {\bibfnamefont {J.}~\bibnamefont {Zhang}}, \bibinfo {author} {\bibfnamefont {K.}~\bibnamefont {Jiang}}, \bibinfo {author} {\bibfnamefont {R.}~\bibnamefont {Zhong}}, \bibinfo {author} {\bibfnamefont {J.}~\bibnamefont {Schneeloch}}, \bibinfo {author} {\bibfnamefont {G.}~\bibnamefont {Gu}}, \bibinfo {author} {\bibfnamefont {L.}~\bibnamefont {Gu}}, \bibinfo {author} {\bibfnamefont {X.}~\bibnamefont {Ma}}, \bibinfo {author} {\bibfnamefont {D.}~\bibnamefont {Zhang}},\ and\ \bibinfo {author} {\bibfnamefont
  {Q.-K.}\ \bibnamefont {Xue}},\ }\bibfield  {title} {\bibinfo {title} {Presence of $s$-wave pairing in josephson junctions made of twisted ultrathin ${\mathrm{bi}}_{2}{\mathrm{sr}}_{2}{\mathrm{cacu}}_{2}{\mathrm{o}}_{8+x}$ flakes},\ }\href {https://doi.org/10.1103/PhysRevX.11.031011} {\bibfield  {journal} {\bibinfo  {journal} {Phys. Rev. X}\ }\textbf {\bibinfo {volume} {11}},\ \bibinfo {pages} {031011} (\bibinfo {year} {2021})}\BibitemShut {NoStop}%
\bibitem [{\citenamefont {Wang}\ \emph {et~al.}(2023)\citenamefont {Wang}, \citenamefont {Zhu}, \citenamefont {Bai}, \citenamefont {Wang}, \citenamefont {Hu}, \citenamefont {Xie}, \citenamefont {Hu}, \citenamefont {Cui}, \citenamefont {Huang}, \citenamefont {Chen}, \citenamefont {Ding}, \citenamefont {Zhao}, \citenamefont {Li}, \citenamefont {Zhang}, \citenamefont {Gu}, \citenamefont {Zhou}, \citenamefont {Zhu}, \citenamefont {Zhang},\ and\ \citenamefont {Xue}}]{Xue2023}%
  \BibitemOpen
  \bibfield  {author} {\bibinfo {author} {\bibfnamefont {H.}~\bibnamefont {Wang}}, \bibinfo {author} {\bibfnamefont {Y.}~\bibnamefont {Zhu}}, \bibinfo {author} {\bibfnamefont {Z.}~\bibnamefont {Bai}}, \bibinfo {author} {\bibfnamefont {Z.}~\bibnamefont {Wang}}, \bibinfo {author} {\bibfnamefont {S.}~\bibnamefont {Hu}}, \bibinfo {author} {\bibfnamefont {H.-Y.}\ \bibnamefont {Xie}}, \bibinfo {author} {\bibfnamefont {X.}~\bibnamefont {Hu}}, \bibinfo {author} {\bibfnamefont {J.}~\bibnamefont {Cui}}, \bibinfo {author} {\bibfnamefont {M.}~\bibnamefont {Huang}}, \bibinfo {author} {\bibfnamefont {J.}~\bibnamefont {Chen}}, \bibinfo {author} {\bibfnamefont {Y.}~\bibnamefont {Ding}}, \bibinfo {author} {\bibfnamefont {L.}~\bibnamefont {Zhao}}, \bibinfo {author} {\bibfnamefont {X.}~\bibnamefont {Li}}, \bibinfo {author} {\bibfnamefont {Q.}~\bibnamefont {Zhang}}, \bibinfo {author} {\bibfnamefont {L.}~\bibnamefont {Gu}}, \bibinfo {author} {\bibfnamefont {X.~J.}\ \bibnamefont {Zhou}}, \bibinfo {author} {\bibfnamefont
  {J.}~\bibnamefont {Zhu}}, \bibinfo {author} {\bibfnamefont {D.}~\bibnamefont {Zhang}},\ and\ \bibinfo {author} {\bibfnamefont {Q.-K.}\ \bibnamefont {Xue}},\ }\bibfield  {title} {\bibinfo {title} {Prominent josephson tunneling between twisted single copper oxide planes of bi2sr2-xlaxcuo6+y},\ }\href {https://doi.org/10.1038/s41467-023-40525-1} {\bibfield  {journal} {\bibinfo  {journal} {Nature Communications}\ }\textbf {\bibinfo {volume} {14}},\ \bibinfo {pages} {5201} (\bibinfo {year} {2023})}\BibitemShut {NoStop}%
\bibitem [{\citenamefont {Martini}\ \emph {et~al.}(2023)\citenamefont {Martini}, \citenamefont {Lee}, \citenamefont {Confalone}, \citenamefont {Shokri}, \citenamefont {Saggau}, \citenamefont {Wolf}, \citenamefont {Gu}, \citenamefont {Watanabe}, \citenamefont {Taniguchi}, \citenamefont {Montemurro}, \citenamefont {Vinokur}, \citenamefont {Nielsch},\ and\ \citenamefont {Poccia}}]{Martini2024}%
  \BibitemOpen
  \bibfield  {author} {\bibinfo {author} {\bibfnamefont {M.}~\bibnamefont {Martini}}, \bibinfo {author} {\bibfnamefont {Y.}~\bibnamefont {Lee}}, \bibinfo {author} {\bibfnamefont {T.}~\bibnamefont {Confalone}}, \bibinfo {author} {\bibfnamefont {S.}~\bibnamefont {Shokri}}, \bibinfo {author} {\bibfnamefont {C.~N.}\ \bibnamefont {Saggau}}, \bibinfo {author} {\bibfnamefont {D.}~\bibnamefont {Wolf}}, \bibinfo {author} {\bibfnamefont {G.}~\bibnamefont {Gu}}, \bibinfo {author} {\bibfnamefont {K.}~\bibnamefont {Watanabe}}, \bibinfo {author} {\bibfnamefont {T.}~\bibnamefont {Taniguchi}}, \bibinfo {author} {\bibfnamefont {D.}~\bibnamefont {Montemurro}}, \bibinfo {author} {\bibfnamefont {V.~M.}\ \bibnamefont {Vinokur}}, \bibinfo {author} {\bibfnamefont {K.}~\bibnamefont {Nielsch}},\ and\ \bibinfo {author} {\bibfnamefont {N.}~\bibnamefont {Poccia}},\ }\bibfield  {title} {\bibinfo {title} {Twisted cuprate van der waals heterostructures with controlled josephson coupling},\ }\href
  {https://doi.org/https://doi.org/10.1016/j.mattod.2023.06.007} {\bibfield  {journal} {\bibinfo  {journal} {Materials Today}\ }\textbf {\bibinfo {volume} {67}},\ \bibinfo {pages} {106} (\bibinfo {year} {2023})}\BibitemShut {NoStop}%
\bibitem [{\citenamefont {Song}\ \emph {et~al.}(2022)\citenamefont {Song}, \citenamefont {Zhang},\ and\ \citenamefont {Vishwanath}}]{Song2022}%
  \BibitemOpen
  \bibfield  {author} {\bibinfo {author} {\bibfnamefont {X.-Y.}\ \bibnamefont {Song}}, \bibinfo {author} {\bibfnamefont {Y.-H.}\ \bibnamefont {Zhang}},\ and\ \bibinfo {author} {\bibfnamefont {A.}~\bibnamefont {Vishwanath}},\ }\bibfield  {title} {\bibinfo {title} {Doping a moir\'e mott insulator: A $t\ensuremath{-}j$ model study of twisted cuprates},\ }\href {https://doi.org/10.1103/PhysRevB.105.L201102} {\bibfield  {journal} {\bibinfo  {journal} {Phys. Rev. B}\ }\textbf {\bibinfo {volume} {105}},\ \bibinfo {pages} {L201102} (\bibinfo {year} {2022})}\BibitemShut {NoStop}%
\bibitem [{\citenamefont {Lu}\ and\ \citenamefont {S\'en\'echal}(2022)}]{Lu2022}%
  \BibitemOpen
  \bibfield  {author} {\bibinfo {author} {\bibfnamefont {X.}~\bibnamefont {Lu}}\ and\ \bibinfo {author} {\bibfnamefont {D.}~\bibnamefont {S\'en\'echal}},\ }\bibfield  {title} {\bibinfo {title} {Doping phase diagram of a hubbard model for twisted bilayer cuprates},\ }\href {https://doi.org/10.1103/PhysRevB.105.245127} {\bibfield  {journal} {\bibinfo  {journal} {Phys. Rev. B}\ }\textbf {\bibinfo {volume} {105}},\ \bibinfo {pages} {245127} (\bibinfo {year} {2022})}\BibitemShut {NoStop}%
\bibitem [{\citenamefont {Fidrysiak}\ \emph {et~al.}(2023)\citenamefont {Fidrysiak}, \citenamefont {Rzeszotarski},\ and\ \citenamefont {Spa\l{}ek}}]{Spalek2023}%
  \BibitemOpen
  \bibfield  {author} {\bibinfo {author} {\bibfnamefont {M.}~\bibnamefont {Fidrysiak}}, \bibinfo {author} {\bibfnamefont {B.}~\bibnamefont {Rzeszotarski}},\ and\ \bibinfo {author} {\bibfnamefont {J.}~\bibnamefont {Spa\l{}ek}},\ }\bibfield  {title} {\bibinfo {title} {Tuning topological superconductivity within the $t\text{\ensuremath{-}}j\text{\ensuremath{-}}u$ model of twisted bilayer cuprates},\ }\href {https://doi.org/10.1103/PhysRevB.108.224509} {\bibfield  {journal} {\bibinfo  {journal} {Phys. Rev. B}\ }\textbf {\bibinfo {volume} {108}},\ \bibinfo {pages} {224509} (\bibinfo {year} {2023})}\BibitemShut {NoStop}%
\bibitem [{\citenamefont {Volkov}\ \emph {et~al.}(2023{\natexlab{b}})\citenamefont {Volkov}, \citenamefont {Wilson}, \citenamefont {Lucht},\ and\ \citenamefont {Pixley}}]{Pixley2023}%
  \BibitemOpen
  \bibfield  {author} {\bibinfo {author} {\bibfnamefont {P.~A.}\ \bibnamefont {Volkov}}, \bibinfo {author} {\bibfnamefont {J.~H.}\ \bibnamefont {Wilson}}, \bibinfo {author} {\bibfnamefont {K.~P.}\ \bibnamefont {Lucht}},\ and\ \bibinfo {author} {\bibfnamefont {J.~H.}\ \bibnamefont {Pixley}},\ }\bibfield  {title} {\bibinfo {title} {Magic angles and correlations in twisted nodal superconductors},\ }\href {https://doi.org/10.1103/PhysRevB.107.174506} {\bibfield  {journal} {\bibinfo  {journal} {Phys. Rev. B}\ }\textbf {\bibinfo {volume} {107}},\ \bibinfo {pages} {174506} (\bibinfo {year} {2023}{\natexlab{b}})}\BibitemShut {NoStop}%
\bibitem [{\citenamefont {B\'elanger}\ and\ \citenamefont {S\'en\'echal}(2024{\natexlab{a}})}]{Senechal2024a}%
  \BibitemOpen
  \bibfield  {author} {\bibinfo {author} {\bibfnamefont {M.}~\bibnamefont {B\'elanger}}\ and\ \bibinfo {author} {\bibfnamefont {D.}~\bibnamefont {S\'en\'echal}},\ }\bibfield  {title} {\bibinfo {title} {Interlayer bias effect on time-reversal symmetry breaking in twisted bilayer cuprates},\ }\href {https://doi.org/10.1103/PhysRevB.109.075111} {\bibfield  {journal} {\bibinfo  {journal} {Phys. Rev. B}\ }\textbf {\bibinfo {volume} {109}},\ \bibinfo {pages} {075111} (\bibinfo {year} {2024}{\natexlab{a}})}\BibitemShut {NoStop}%
\bibitem [{\citenamefont {B\'elanger}\ and\ \citenamefont {S\'en\'echal}(2024{\natexlab{b}})}]{Senechal2024b}%
  \BibitemOpen
  \bibfield  {author} {\bibinfo {author} {\bibfnamefont {M.}~\bibnamefont {B\'elanger}}\ and\ \bibinfo {author} {\bibfnamefont {D.}~\bibnamefont {S\'en\'echal}},\ }\bibfield  {title} {\bibinfo {title} {Doping dependence of chiral superconductivity in near ${45}^{\ensuremath{\circ}}$ twisted bilayer cuprates},\ }\href {https://doi.org/10.1103/PhysRevB.109.045111} {\bibfield  {journal} {\bibinfo  {journal} {Phys. Rev. B}\ }\textbf {\bibinfo {volume} {109}},\ \bibinfo {pages} {045111} (\bibinfo {year} {2024}{\natexlab{b}})}\BibitemShut {NoStop}%
\bibitem [{\citenamefont {Haenel}\ \emph {et~al.}(2022)\citenamefont {Haenel}, \citenamefont {Tummuru},\ and\ \citenamefont {Franz}}]{Haenel2022}%
  \BibitemOpen
  \bibfield  {author} {\bibinfo {author} {\bibfnamefont {R.}~\bibnamefont {Haenel}}, \bibinfo {author} {\bibfnamefont {T.}~\bibnamefont {Tummuru}},\ and\ \bibinfo {author} {\bibfnamefont {M.}~\bibnamefont {Franz}},\ }\bibfield  {title} {\bibinfo {title} {Incoherent tunneling and topological superconductivity in twisted cuprate bilayers},\ }\href {https://doi.org/10.1103/PhysRevB.106.104505} {\bibfield  {journal} {\bibinfo  {journal} {Phys. Rev. B}\ }\textbf {\bibinfo {volume} {106}},\ \bibinfo {pages} {104505} (\bibinfo {year} {2022})}\BibitemShut {NoStop}%
\bibitem [{\citenamefont {Yuan}\ \emph {et~al.}(2023)\citenamefont {Yuan}, \citenamefont {Vituri}, \citenamefont {Berg}, \citenamefont {Spivak},\ and\ \citenamefont {Kivelson}}]{Yuan2023}%
  \BibitemOpen
  \bibfield  {author} {\bibinfo {author} {\bibfnamefont {A.~C.}\ \bibnamefont {Yuan}}, \bibinfo {author} {\bibfnamefont {Y.}~\bibnamefont {Vituri}}, \bibinfo {author} {\bibfnamefont {E.}~\bibnamefont {Berg}}, \bibinfo {author} {\bibfnamefont {B.}~\bibnamefont {Spivak}},\ and\ \bibinfo {author} {\bibfnamefont {S.~A.}\ \bibnamefont {Kivelson}},\ }\bibfield  {title} {\bibinfo {title} {Inhomogeneity-induced time-reversal symmetry breaking in cuprate twist junctions},\ }\href {https://doi.org/10.1103/PhysRevB.108.L100505} {\bibfield  {journal} {\bibinfo  {journal} {Phys. Rev. B}\ }\textbf {\bibinfo {volume} {108}},\ \bibinfo {pages} {L100505} (\bibinfo {year} {2023})}\BibitemShut {NoStop}%
\bibitem [{\citenamefont {Volkov}\ \emph {et~al.}(2023{\natexlab{c}})\citenamefont {Volkov}, \citenamefont {Wilson}, \citenamefont {Lucht},\ and\ \citenamefont {Pixley}}]{Volkov2023}%
  \BibitemOpen
  \bibfield  {author} {\bibinfo {author} {\bibfnamefont {P.~A.}\ \bibnamefont {Volkov}}, \bibinfo {author} {\bibfnamefont {J.~H.}\ \bibnamefont {Wilson}}, \bibinfo {author} {\bibfnamefont {K.~P.}\ \bibnamefont {Lucht}},\ and\ \bibinfo {author} {\bibfnamefont {J.~H.}\ \bibnamefont {Pixley}},\ }\bibfield  {title} {\bibinfo {title} {Current- and field-induced topology in twisted nodal superconductors},\ }\href {https://doi.org/10.1103/PhysRevLett.130.186001} {\bibfield  {journal} {\bibinfo  {journal} {Phys. Rev. Lett.}\ }\textbf {\bibinfo {volume} {130}},\ \bibinfo {pages} {186001} (\bibinfo {year} {2023}{\natexlab{c}})}\BibitemShut {NoStop}%
\bibitem [{\citenamefont {Can}\ \emph {et~al.}(2021{\natexlab{b}})\citenamefont {Can}, \citenamefont {Zhang}, \citenamefont {Kallin},\ and\ \citenamefont {Franz}}]{Can2021_2}%
  \BibitemOpen
  \bibfield  {author} {\bibinfo {author} {\bibfnamefont {O.}~\bibnamefont {Can}}, \bibinfo {author} {\bibfnamefont {X.-X.}\ \bibnamefont {Zhang}}, \bibinfo {author} {\bibfnamefont {C.}~\bibnamefont {Kallin}},\ and\ \bibinfo {author} {\bibfnamefont {M.}~\bibnamefont {Franz}},\ }\bibfield  {title} {\bibinfo {title} {Probing time reversal symmetry breaking topological superconductivity in twisted double layer copper oxides with polar kerr effect},\ }\href {https://doi.org/10.1103/PhysRevLett.127.157001} {\bibfield  {journal} {\bibinfo  {journal} {Phys. Rev. Lett.}\ }\textbf {\bibinfo {volume} {127}},\ \bibinfo {pages} {157001} (\bibinfo {year} {2021}{\natexlab{b}})}\BibitemShut {NoStop}%
\bibitem [{\citenamefont {Holmvall}\ \emph {et~al.}(2023)\citenamefont {Holmvall}, \citenamefont {Wall-Wennerdal},\ and\ \citenamefont {Black-Schaffer}}]{Holmvall2023}%
  \BibitemOpen
  \bibfield  {author} {\bibinfo {author} {\bibfnamefont {P.}~\bibnamefont {Holmvall}}, \bibinfo {author} {\bibfnamefont {N.}~\bibnamefont {Wall-Wennerdal}},\ and\ \bibinfo {author} {\bibfnamefont {A.~M.}\ \bibnamefont {Black-Schaffer}},\ }\bibfield  {title} {\bibinfo {title} {Robust and tunable coreless vortices and fractional vortices in chiral $d$-wave superconductors},\ }\href {https://doi.org/10.1103/PhysRevB.108.094511} {\bibfield  {journal} {\bibinfo  {journal} {Phys. Rev. B}\ }\textbf {\bibinfo {volume} {108}},\ \bibinfo {pages} {094511} (\bibinfo {year} {2023})}\BibitemShut {NoStop}%
\bibitem [{\citenamefont {H{\aa}kansson}\ \emph {et~al.}(2015)\citenamefont {H{\aa}kansson}, \citenamefont {L{\"o}fwander},\ and\ \citenamefont {Fogelstr{\"o}m}}]{Hakansson2015}%
  \BibitemOpen
  \bibfield  {author} {\bibinfo {author} {\bibfnamefont {M.}~\bibnamefont {H{\aa}kansson}}, \bibinfo {author} {\bibfnamefont {T.}~\bibnamefont {L{\"o}fwander}},\ and\ \bibinfo {author} {\bibfnamefont {M.}~\bibnamefont {Fogelstr{\"o}m}},\ }\bibfield  {title} {\bibinfo {title} {Spontaneously broken time-reversal symmetry in high-temperature superconductors},\ }\href {https://doi.org/10.1038/nphys3383} {\bibfield  {journal} {\bibinfo  {journal} {Nature Physics}\ }\textbf {\bibinfo {volume} {11}},\ \bibinfo {pages} {755} (\bibinfo {year} {2015})}\BibitemShut {NoStop}%
\bibitem [{\citenamefont {Holmvall}\ \emph {et~al.}(2018)\citenamefont {Holmvall}, \citenamefont {Vorontsov}, \citenamefont {Fogelstr{\"o}m},\ and\ \citenamefont {L{\"o}fwander}}]{Holmvall2018}%
  \BibitemOpen
  \bibfield  {author} {\bibinfo {author} {\bibfnamefont {P.}~\bibnamefont {Holmvall}}, \bibinfo {author} {\bibfnamefont {A.~B.}\ \bibnamefont {Vorontsov}}, \bibinfo {author} {\bibfnamefont {M.}~\bibnamefont {Fogelstr{\"o}m}},\ and\ \bibinfo {author} {\bibfnamefont {T.}~\bibnamefont {L{\"o}fwander}},\ }\bibfield  {title} {\bibinfo {title} {Broken translational symmetry at edges of high-temperature superconductors},\ }\href {https://doi.org/10.1038/s41467-018-04531-y} {\bibfield  {journal} {\bibinfo  {journal} {Nature Communications}\ }\textbf {\bibinfo {volume} {9}},\ \bibinfo {pages} {2190} (\bibinfo {year} {2018})}\BibitemShut {NoStop}%
\bibitem [{\citenamefont {Holmvall}\ \emph {et~al.}(2019)\citenamefont {Holmvall}, \citenamefont {Vorontsov}, \citenamefont {Fogelstr\"om},\ and\ \citenamefont {L\"ofwander}}]{Holmvall2019}%
  \BibitemOpen
  \bibfield  {author} {\bibinfo {author} {\bibfnamefont {P.}~\bibnamefont {Holmvall}}, \bibinfo {author} {\bibfnamefont {A.~B.}\ \bibnamefont {Vorontsov}}, \bibinfo {author} {\bibfnamefont {M.}~\bibnamefont {Fogelstr\"om}},\ and\ \bibinfo {author} {\bibfnamefont {T.}~\bibnamefont {L\"ofwander}},\ }\bibfield  {title} {\bibinfo {title} {Spontaneous symmetry breaking at surfaces of $d$-wave superconductors: Influence of geometry and surface ruggedness},\ }\href {https://doi.org/10.1103/PhysRevB.99.184511} {\bibfield  {journal} {\bibinfo  {journal} {Phys. Rev. B}\ }\textbf {\bibinfo {volume} {99}},\ \bibinfo {pages} {184511} (\bibinfo {year} {2019})}\BibitemShut {NoStop}%
\bibitem [{\citenamefont {Wennerdal}\ \emph {et~al.}(2020)\citenamefont {Wennerdal}, \citenamefont {Ask}, \citenamefont {Holmvall}, \citenamefont {L\"ofwander},\ and\ \citenamefont {Fogelstr\"om}}]{Wennerdal2020}%
  \BibitemOpen
  \bibfield  {author} {\bibinfo {author} {\bibfnamefont {N.~W.}\ \bibnamefont {Wennerdal}}, \bibinfo {author} {\bibfnamefont {A.}~\bibnamefont {Ask}}, \bibinfo {author} {\bibfnamefont {P.}~\bibnamefont {Holmvall}}, \bibinfo {author} {\bibfnamefont {T.}~\bibnamefont {L\"ofwander}},\ and\ \bibinfo {author} {\bibfnamefont {M.}~\bibnamefont {Fogelstr\"om}},\ }\bibfield  {title} {\bibinfo {title} {Breaking time-reversal and translational symmetry at edges of $d$-wave superconductors: Microscopic theory and comparison with quasiclassical theory},\ }\href {https://doi.org/10.1103/PhysRevResearch.2.043198} {\bibfield  {journal} {\bibinfo  {journal} {Phys. Rev. Res.}\ }\textbf {\bibinfo {volume} {2}},\ \bibinfo {pages} {043198} (\bibinfo {year} {2020})}\BibitemShut {NoStop}%
\bibitem [{\citenamefont {Tummuru}\ \emph {et~al.}(2022)\citenamefont {Tummuru}, \citenamefont {Plugge},\ and\ \citenamefont {Franz}}]{Tummuru2022}%
  \BibitemOpen
  \bibfield  {author} {\bibinfo {author} {\bibfnamefont {T.}~\bibnamefont {Tummuru}}, \bibinfo {author} {\bibfnamefont {S.}~\bibnamefont {Plugge}},\ and\ \bibinfo {author} {\bibfnamefont {M.}~\bibnamefont {Franz}},\ }\bibfield  {title} {\bibinfo {title} {Josephson effects in twisted cuprate bilayers},\ }\href {https://doi.org/10.1103/PhysRevB.105.064501} {\bibfield  {journal} {\bibinfo  {journal} {Phys. Rev. B}\ }\textbf {\bibinfo {volume} {105}},\ \bibinfo {pages} {064501} (\bibinfo {year} {2022})}\BibitemShut {NoStop}%
\bibitem [{\citenamefont {Nie}\ \emph {et~al.}(2020)\citenamefont {Nie}, \citenamefont {Huang},\ and\ \citenamefont {Yao}}]{Nie2020}%
  \BibitemOpen
  \bibfield  {author} {\bibinfo {author} {\bibfnamefont {W.}~\bibnamefont {Nie}}, \bibinfo {author} {\bibfnamefont {W.}~\bibnamefont {Huang}},\ and\ \bibinfo {author} {\bibfnamefont {H.}~\bibnamefont {Yao}},\ }\bibfield  {title} {\bibinfo {title} {Edge current and orbital angular momentum of chiral superfluids revisited},\ }\href {https://doi.org/10.1103/PhysRevB.102.054502} {\bibfield  {journal} {\bibinfo  {journal} {Phys. Rev. B}\ }\textbf {\bibinfo {volume} {102}},\ \bibinfo {pages} {054502} (\bibinfo {year} {2020})}\BibitemShut {NoStop}%
\bibitem [{\citenamefont {Tummuru}\ \emph {et~al.}(2021)\citenamefont {Tummuru}, \citenamefont {Can},\ and\ \citenamefont {Franz}}]{Tummuru2021}%
  \BibitemOpen
  \bibfield  {author} {\bibinfo {author} {\bibfnamefont {T.}~\bibnamefont {Tummuru}}, \bibinfo {author} {\bibfnamefont {O.}~\bibnamefont {Can}},\ and\ \bibinfo {author} {\bibfnamefont {M.}~\bibnamefont {Franz}},\ }\bibfield  {title} {\bibinfo {title} {Chiral $p$-wave superconductivity in a twisted array of proximitized quantum wires},\ }\href {https://doi.org/10.1103/PhysRevB.103.L100501} {\bibfield  {journal} {\bibinfo  {journal} {Phys. Rev. B}\ }\textbf {\bibinfo {volume} {103}},\ \bibinfo {pages} {L100501} (\bibinfo {year} {2021})}\BibitemShut {NoStop}%
\bibitem [{\citenamefont {Hartmann}(1999)}]{Hartmann1999}%
  \BibitemOpen
  \bibfield  {author} {\bibinfo {author} {\bibfnamefont {U.}~\bibnamefont {Hartmann}},\ }\bibfield  {title} {\bibinfo {title} {Magnetic force microscopy},\ }\href {https://doi.org/10.1146/annurev.matsci.29.1.53} {\bibfield  {journal} {\bibinfo  {journal} {Annual Review of Materials Science}\ }\textbf {\bibinfo {volume} {29}},\ \bibinfo {pages} {53} (\bibinfo {year} {1999})}\BibitemShut {NoStop}%
\bibitem [{\citenamefont {Casola}\ \emph {et~al.}(2018)\citenamefont {Casola}, \citenamefont {van~der Sar},\ and\ \citenamefont {Yacoby}}]{Casola2018}%
  \BibitemOpen
  \bibfield  {author} {\bibinfo {author} {\bibfnamefont {F.}~\bibnamefont {Casola}}, \bibinfo {author} {\bibfnamefont {T.}~\bibnamefont {van~der Sar}},\ and\ \bibinfo {author} {\bibfnamefont {A.}~\bibnamefont {Yacoby}},\ }\bibfield  {title} {\bibinfo {title} {Probing condensed matter physics with magnetometry based on nitrogen-vacancy centres in diamond},\ }\href {https://doi.org/10.1038/natrevmats.2017.88} {\bibfield  {journal} {\bibinfo  {journal} {Nature Reviews Materials}\ }\textbf {\bibinfo {volume} {3}},\ \bibinfo {pages} {17088} (\bibinfo {year} {2018})}\BibitemShut {NoStop}%
\bibitem [{\citenamefont {Kirtley}(2010)}]{Kirtley2010}%
  \BibitemOpen
  \bibfield  {author} {\bibinfo {author} {\bibfnamefont {J.~R.}\ \bibnamefont {Kirtley}},\ }\bibfield  {title} {\bibinfo {title} {Fundamental studies of superconductors using scanning magnetic imaging},\ }\href {https://doi.org/10.1088/0034-4885/73/12/126501} {\bibfield  {journal} {\bibinfo  {journal} {Reports on Progress in Physics}\ }\textbf {\bibinfo {volume} {73}},\ \bibinfo {pages} {126501} (\bibinfo {year} {2010})}\BibitemShut {NoStop}%
\bibitem [{\citenamefont {Tsuei}\ \emph {et~al.}(1994)\citenamefont {Tsuei}, \citenamefont {Kirtley}, \citenamefont {Chi}, \citenamefont {Yu-Jahnes}, \citenamefont {Gupta}, \citenamefont {Shaw}, \citenamefont {Sun},\ and\ \citenamefont {Ketchen}}]{Tsuei1994}%
  \BibitemOpen
  \bibfield  {author} {\bibinfo {author} {\bibfnamefont {C.~C.}\ \bibnamefont {Tsuei}}, \bibinfo {author} {\bibfnamefont {J.~R.}\ \bibnamefont {Kirtley}}, \bibinfo {author} {\bibfnamefont {C.~C.}\ \bibnamefont {Chi}}, \bibinfo {author} {\bibfnamefont {L.~S.}\ \bibnamefont {Yu-Jahnes}}, \bibinfo {author} {\bibfnamefont {A.}~\bibnamefont {Gupta}}, \bibinfo {author} {\bibfnamefont {T.}~\bibnamefont {Shaw}}, \bibinfo {author} {\bibfnamefont {J.~Z.}\ \bibnamefont {Sun}},\ and\ \bibinfo {author} {\bibfnamefont {M.~B.}\ \bibnamefont {Ketchen}},\ }\bibfield  {title} {\bibinfo {title} {Pairing symmetry and flux quantization in a tricrystal superconducting ring of $\mathrm{Y}{\mathrm{ba}}_{2}{\mathrm{cu}}_{3}{\mathrm{o}}_{7\ensuremath{-}\ensuremath{\delta}}$},\ }\href {https://doi.org/10.1103/PhysRevLett.73.593} {\bibfield  {journal} {\bibinfo  {journal} {Phys. Rev. Lett.}\ }\textbf {\bibinfo {volume} {73}},\ \bibinfo {pages} {593} (\bibinfo {year} {1994})}\BibitemShut {NoStop}%
\bibitem [{\citenamefont {Kalisky}\ \emph {et~al.}(2011)\citenamefont {Kalisky}, \citenamefont {Kirtley}, \citenamefont {Analytis}, \citenamefont {Chu}, \citenamefont {Fisher},\ and\ \citenamefont {Moler}}]{Kalisky2011}%
  \BibitemOpen
  \bibfield  {author} {\bibinfo {author} {\bibfnamefont {B.}~\bibnamefont {Kalisky}}, \bibinfo {author} {\bibfnamefont {J.~R.}\ \bibnamefont {Kirtley}}, \bibinfo {author} {\bibfnamefont {J.~G.}\ \bibnamefont {Analytis}}, \bibinfo {author} {\bibfnamefont {J.-H.}\ \bibnamefont {Chu}}, \bibinfo {author} {\bibfnamefont {I.~R.}\ \bibnamefont {Fisher}},\ and\ \bibinfo {author} {\bibfnamefont {K.~A.}\ \bibnamefont {Moler}},\ }\bibfield  {title} {\bibinfo {title} {Behavior of vortices near twin boundaries in underdoped ba(fe${}_{1\ensuremath{-}x}{\mathrm{co}}_{x}){}_{2}{\mathrm{as}}_{2}$},\ }\href {https://doi.org/10.1103/PhysRevB.83.064511} {\bibfield  {journal} {\bibinfo  {journal} {Phys. Rev. B}\ }\textbf {\bibinfo {volume} {83}},\ \bibinfo {pages} {064511} (\bibinfo {year} {2011})}\BibitemShut {NoStop}%
\bibitem [{\citenamefont {Christensen}\ \emph {et~al.}(2019)\citenamefont {Christensen}, \citenamefont {Frenkel}, \citenamefont {Chen}, \citenamefont {Xie}, \citenamefont {Chen}, \citenamefont {Hikita}, \citenamefont {Smith}, \citenamefont {Klein}, \citenamefont {Hwang}, \citenamefont {Pryds},\ and\ \citenamefont {Kalisky}}]{Christensen2019}%
  \BibitemOpen
  \bibfield  {author} {\bibinfo {author} {\bibfnamefont {D.~V.}\ \bibnamefont {Christensen}}, \bibinfo {author} {\bibfnamefont {Y.}~\bibnamefont {Frenkel}}, \bibinfo {author} {\bibfnamefont {Y.~Z.}\ \bibnamefont {Chen}}, \bibinfo {author} {\bibfnamefont {Y.~W.}\ \bibnamefont {Xie}}, \bibinfo {author} {\bibfnamefont {Z.~Y.}\ \bibnamefont {Chen}}, \bibinfo {author} {\bibfnamefont {Y.}~\bibnamefont {Hikita}}, \bibinfo {author} {\bibfnamefont {A.}~\bibnamefont {Smith}}, \bibinfo {author} {\bibfnamefont {L.}~\bibnamefont {Klein}}, \bibinfo {author} {\bibfnamefont {H.~Y.}\ \bibnamefont {Hwang}}, \bibinfo {author} {\bibfnamefont {N.}~\bibnamefont {Pryds}},\ and\ \bibinfo {author} {\bibfnamefont {B.}~\bibnamefont {Kalisky}},\ }\bibfield  {title} {\bibinfo {title} {Strain-tunable magnetism at oxide domain walls},\ }\href {https://doi.org/10.1038/s41567-018-0363-x} {\bibfield  {journal} {\bibinfo  {journal} {Nature Physics}\ }\textbf {\bibinfo {volume} {15}},\ \bibinfo {pages} {269} (\bibinfo {year}
  {2019})}\BibitemShut {NoStop}%
\bibitem [{\citenamefont {Uri}\ \emph {et~al.}(2020)\citenamefont {Uri}, \citenamefont {Kim}, \citenamefont {Bagani}, \citenamefont {Lewandowski}, \citenamefont {Grover}, \citenamefont {Auerbach}, \citenamefont {Lachman}, \citenamefont {Myasoedov}, \citenamefont {Taniguchi}, \citenamefont {Watanabe}, \citenamefont {Smet},\ and\ \citenamefont {Zeldov}}]{Uri2020}%
  \BibitemOpen
  \bibfield  {author} {\bibinfo {author} {\bibfnamefont {A.}~\bibnamefont {Uri}}, \bibinfo {author} {\bibfnamefont {Y.}~\bibnamefont {Kim}}, \bibinfo {author} {\bibfnamefont {K.}~\bibnamefont {Bagani}}, \bibinfo {author} {\bibfnamefont {C.~K.}\ \bibnamefont {Lewandowski}}, \bibinfo {author} {\bibfnamefont {S.}~\bibnamefont {Grover}}, \bibinfo {author} {\bibfnamefont {N.}~\bibnamefont {Auerbach}}, \bibinfo {author} {\bibfnamefont {E.~O.}\ \bibnamefont {Lachman}}, \bibinfo {author} {\bibfnamefont {Y.}~\bibnamefont {Myasoedov}}, \bibinfo {author} {\bibfnamefont {T.}~\bibnamefont {Taniguchi}}, \bibinfo {author} {\bibfnamefont {K.}~\bibnamefont {Watanabe}}, \bibinfo {author} {\bibfnamefont {J.}~\bibnamefont {Smet}},\ and\ \bibinfo {author} {\bibfnamefont {E.}~\bibnamefont {Zeldov}},\ }\bibfield  {title} {\bibinfo {title} {{Nanoscale imaging of equilibrium quantum Hall edge currents and of the magnetic monopole response in graphene}},\ }\href {https://doi.org/10.1038/s41567-019-0713-3} {\bibfield  {journal}
  {\bibinfo  {journal} {Nature Physics}\ }\textbf {\bibinfo {volume} {16}},\ \bibinfo {pages} {164} (\bibinfo {year} {2020})}\BibitemShut {NoStop}%
\bibitem [{\citenamefont {Persky}\ \emph {et~al.}(2022)\citenamefont {Persky}, \citenamefont {Sochnikov},\ and\ \citenamefont {Kalisky}}]{Persky2021}%
  \BibitemOpen
  \bibfield  {author} {\bibinfo {author} {\bibfnamefont {E.}~\bibnamefont {Persky}}, \bibinfo {author} {\bibfnamefont {I.}~\bibnamefont {Sochnikov}},\ and\ \bibinfo {author} {\bibfnamefont {B.}~\bibnamefont {Kalisky}},\ }\bibfield  {title} {\bibinfo {title} {Studying quantum materials with scanning squid microscopy},\ }\href {https://doi.org/10.1146/annurev-conmatphys-031620-104226} {\bibfield  {journal} {\bibinfo  {journal} {Annual Review of Condensed Matter Physics}\ }\textbf {\bibinfo {volume} {13}},\ \bibinfo {pages} {385} (\bibinfo {year} {2022})}\BibitemShut {NoStop}%
\bibitem [{\citenamefont {Kirtley}\ \emph {et~al.}(2016)\citenamefont {Kirtley}, \citenamefont {Paulius}, \citenamefont {Rosenberg}, \citenamefont {Palmstrom}, \citenamefont {Holland}, \citenamefont {Spanton}, \citenamefont {Schiessl}, \citenamefont {Jermain}, \citenamefont {Gibbons}, \citenamefont {Fung}, \citenamefont {Huber}, \citenamefont {Ralph}, \citenamefont {Ketchen}, \citenamefont {Gibson},\ and\ \citenamefont {Moler}}]{Kirtley2016}%
  \BibitemOpen
  \bibfield  {author} {\bibinfo {author} {\bibfnamefont {J.~R.}\ \bibnamefont {Kirtley}}, \bibinfo {author} {\bibfnamefont {L.}~\bibnamefont {Paulius}}, \bibinfo {author} {\bibfnamefont {A.~J.}\ \bibnamefont {Rosenberg}}, \bibinfo {author} {\bibfnamefont {J.~C.}\ \bibnamefont {Palmstrom}}, \bibinfo {author} {\bibfnamefont {C.~M.}\ \bibnamefont {Holland}}, \bibinfo {author} {\bibfnamefont {E.~M.}\ \bibnamefont {Spanton}}, \bibinfo {author} {\bibfnamefont {D.}~\bibnamefont {Schiessl}}, \bibinfo {author} {\bibfnamefont {C.~L.}\ \bibnamefont {Jermain}}, \bibinfo {author} {\bibfnamefont {J.}~\bibnamefont {Gibbons}}, \bibinfo {author} {\bibfnamefont {Y.-K.-K.}\ \bibnamefont {Fung}}, \bibinfo {author} {\bibfnamefont {M.~E.}\ \bibnamefont {Huber}}, \bibinfo {author} {\bibfnamefont {D.~C.}\ \bibnamefont {Ralph}}, \bibinfo {author} {\bibfnamefont {M.~B.}\ \bibnamefont {Ketchen}}, \bibinfo {author} {\bibfnamefont {J.}~\bibnamefont {Gibson}, \bibfnamefont {Gerald~W.}},\ and\ \bibinfo {author} {\bibfnamefont {K.~A.}\
  \bibnamefont {Moler}},\ }\bibfield  {title} {\bibinfo {title} {{Scanning SQUID susceptometers with sub-micron spatial resolution}},\ }\href {https://doi.org/10.1063/1.4961982} {\bibfield  {journal} {\bibinfo  {journal} {Review of Scientific Instruments}\ }\textbf {\bibinfo {volume} {87}},\ \bibinfo {pages} {093702} (\bibinfo {year} {2016})}\BibitemShut {NoStop}%
\bibitem [{\citenamefont {Cui}\ \emph {et~al.}(2017)\citenamefont {Cui}, \citenamefont {Kirtley}, \citenamefont {Wang}, \citenamefont {Kratz}, \citenamefont {Rosenberg}, \citenamefont {Watson}, \citenamefont {Gibson}, \citenamefont {Ketchen},\ and\ \citenamefont {Moler}}]{Cui2017}%
  \BibitemOpen
  \bibfield  {author} {\bibinfo {author} {\bibfnamefont {Z.}~\bibnamefont {Cui}}, \bibinfo {author} {\bibfnamefont {J.~R.}\ \bibnamefont {Kirtley}}, \bibinfo {author} {\bibfnamefont {Y.}~\bibnamefont {Wang}}, \bibinfo {author} {\bibfnamefont {P.~A.}\ \bibnamefont {Kratz}}, \bibinfo {author} {\bibfnamefont {A.~J.}\ \bibnamefont {Rosenberg}}, \bibinfo {author} {\bibfnamefont {C.~A.}\ \bibnamefont {Watson}}, \bibinfo {author} {\bibfnamefont {J.}~\bibnamefont {Gibson}, \bibfnamefont {Gerald~W.}}, \bibinfo {author} {\bibfnamefont {M.~B.}\ \bibnamefont {Ketchen}},\ and\ \bibinfo {author} {\bibfnamefont {K.~A.}\ \bibnamefont {Moler}},\ }\bibfield  {title} {\bibinfo {title} {{Scanning SQUID sampler with 40-ps time resolution}},\ }\href {https://doi.org/10.1063/1.4986525} {\bibfield  {journal} {\bibinfo  {journal} {Review of Scientific Instruments}\ }\textbf {\bibinfo {volume} {88}},\ \bibinfo {pages} {083703} (\bibinfo {year} {2017})}\BibitemShut {NoStop}%
\bibitem [{\citenamefont {Vasyukov}\ \emph {et~al.}(2013)\citenamefont {Vasyukov}, \citenamefont {Anahory}, \citenamefont {Embon}, \citenamefont {Halbertal}, \citenamefont {Cuppens}, \citenamefont {Neeman}, \citenamefont {Finkler}, \citenamefont {Segev}, \citenamefont {Myasoedov}, \citenamefont {Rappaport}, \citenamefont {Huber},\ and\ \citenamefont {Zeldov}}]{Vasyukov2013}%
  \BibitemOpen
  \bibfield  {author} {\bibinfo {author} {\bibfnamefont {D.}~\bibnamefont {Vasyukov}}, \bibinfo {author} {\bibfnamefont {Y.}~\bibnamefont {Anahory}}, \bibinfo {author} {\bibfnamefont {L.}~\bibnamefont {Embon}}, \bibinfo {author} {\bibfnamefont {D.}~\bibnamefont {Halbertal}}, \bibinfo {author} {\bibfnamefont {J.}~\bibnamefont {Cuppens}}, \bibinfo {author} {\bibfnamefont {L.}~\bibnamefont {Neeman}}, \bibinfo {author} {\bibfnamefont {A.}~\bibnamefont {Finkler}}, \bibinfo {author} {\bibfnamefont {Y.}~\bibnamefont {Segev}}, \bibinfo {author} {\bibfnamefont {Y.}~\bibnamefont {Myasoedov}}, \bibinfo {author} {\bibfnamefont {M.~L.}\ \bibnamefont {Rappaport}}, \bibinfo {author} {\bibfnamefont {M.~E.}\ \bibnamefont {Huber}},\ and\ \bibinfo {author} {\bibfnamefont {E.}~\bibnamefont {Zeldov}},\ }\bibfield  {title} {\bibinfo {title} {A scanning superconducting quantum interference device with single electron spin sensitivity},\ }\href {https://doi.org/10.1038/nnano.2013.169} {\bibfield  {journal} {\bibinfo  {journal}
  {Nature Nanotechnology}\ }\textbf {\bibinfo {volume} {8}},\ \bibinfo {pages} {639} (\bibinfo {year} {2013})}\BibitemShut {NoStop}%
\bibitem [{\citenamefont {Liu}\ \emph {et~al.}(2023{\natexlab{a}})\citenamefont {Liu}, \citenamefont {Zhou}, \citenamefont {Wu},\ and\ \citenamefont {Yang}}]{Liu2023}%
  \BibitemOpen
  \bibfield  {author} {\bibinfo {author} {\bibfnamefont {Y.-B.}\ \bibnamefont {Liu}}, \bibinfo {author} {\bibfnamefont {J.}~\bibnamefont {Zhou}}, \bibinfo {author} {\bibfnamefont {C.}~\bibnamefont {Wu}},\ and\ \bibinfo {author} {\bibfnamefont {F.}~\bibnamefont {Yang}},\ }\bibfield  {title} {\bibinfo {title} {Charge-4e superconductivity and chiral metal in 45$\,^{\circ}$-twisted bilayer cuprates and related bilayers},\ }\href {https://doi.org/10.1038/s41467-023-43782-2} {\bibfield  {journal} {\bibinfo  {journal} {Nature Communications}\ }\textbf {\bibinfo {volume} {14}},\ \bibinfo {pages} {7926} (\bibinfo {year} {2023}{\natexlab{a}})}\BibitemShut {NoStop}%
\bibitem [{\citenamefont {Li}\ and\ \citenamefont {Liu}(2023)}]{Li2023}%
  \BibitemOpen
  \bibfield  {author} {\bibinfo {author} {\bibfnamefont {Y.-X.}\ \bibnamefont {Li}}\ and\ \bibinfo {author} {\bibfnamefont {C.-C.}\ \bibnamefont {Liu}},\ }\bibfield  {title} {\bibinfo {title} {High-temperature majorana corner modes in a $d+i{d}^{\ensuremath{'}}$ superconductor heterostructure: Application to twisted bilayer cuprate superconductors},\ }\href {https://doi.org/10.1103/PhysRevB.107.235125} {\bibfield  {journal} {\bibinfo  {journal} {Phys. Rev. B}\ }\textbf {\bibinfo {volume} {107}},\ \bibinfo {pages} {235125} (\bibinfo {year} {2023})}\BibitemShut {NoStop}%
\bibitem [{\citenamefont {Margalit}\ \emph {et~al.}(2022)\citenamefont {Margalit}, \citenamefont {Yan}, \citenamefont {Franz},\ and\ \citenamefont {Oreg}}]{Margalit2022}%
  \BibitemOpen
  \bibfield  {author} {\bibinfo {author} {\bibfnamefont {G.}~\bibnamefont {Margalit}}, \bibinfo {author} {\bibfnamefont {B.}~\bibnamefont {Yan}}, \bibinfo {author} {\bibfnamefont {M.}~\bibnamefont {Franz}},\ and\ \bibinfo {author} {\bibfnamefont {Y.}~\bibnamefont {Oreg}},\ }\bibfield  {title} {\bibinfo {title} {Chiral majorana modes via proximity to a twisted cuprate bilayer},\ }\href {https://doi.org/10.1103/PhysRevB.106.205424} {\bibfield  {journal} {\bibinfo  {journal} {Phys. Rev. B}\ }\textbf {\bibinfo {volume} {106}},\ \bibinfo {pages} {205424} (\bibinfo {year} {2022})}\BibitemShut {NoStop}%
\bibitem [{\citenamefont {Mercado}\ \emph {et~al.}(2022)\citenamefont {Mercado}, \citenamefont {Sahoo},\ and\ \citenamefont {Franz}}]{Mercado2022}%
  \BibitemOpen
  \bibfield  {author} {\bibinfo {author} {\bibfnamefont {A.}~\bibnamefont {Mercado}}, \bibinfo {author} {\bibfnamefont {S.}~\bibnamefont {Sahoo}},\ and\ \bibinfo {author} {\bibfnamefont {M.}~\bibnamefont {Franz}},\ }\bibfield  {title} {\bibinfo {title} {High-temperature majorana zero modes},\ }\href {https://doi.org/10.1103/PhysRevLett.128.137002} {\bibfield  {journal} {\bibinfo  {journal} {Phys. Rev. Lett.}\ }\textbf {\bibinfo {volume} {128}},\ \bibinfo {pages} {137002} (\bibinfo {year} {2022})}\BibitemShut {NoStop}%
\bibitem [{\citenamefont {Eugenio}\ and\ \citenamefont {Vafek}(2023)}]{Vafek2023}%
  \BibitemOpen
  \bibfield  {author} {\bibinfo {author} {\bibfnamefont {P.~M.}\ \bibnamefont {Eugenio}}\ and\ \bibinfo {author} {\bibfnamefont {O.}~\bibnamefont {Vafek}},\ }\bibfield  {title} {\bibinfo {title} {{Twisted-bilayer FeSe and the Fe-based superlattices}},\ }\href {https://doi.org/10.21468/SciPostPhys.15.3.081} {\bibfield  {journal} {\bibinfo  {journal} {SciPost Phys.}\ }\textbf {\bibinfo {volume} {15}},\ \bibinfo {pages} {081} (\bibinfo {year} {2023})}\BibitemShut {NoStop}%
\bibitem [{\citenamefont {Zhou}\ \emph {et~al.}(2023)\citenamefont {Zhou}, \citenamefont {Egan}, \citenamefont {Kush},\ and\ \citenamefont {Franz}}]{Benjamin2023}%
  \BibitemOpen
  \bibfield  {author} {\bibinfo {author} {\bibfnamefont {B.~T.}\ \bibnamefont {Zhou}}, \bibinfo {author} {\bibfnamefont {S.}~\bibnamefont {Egan}}, \bibinfo {author} {\bibfnamefont {D.}~\bibnamefont {Kush}},\ and\ \bibinfo {author} {\bibfnamefont {M.}~\bibnamefont {Franz}},\ }\bibfield  {title} {\bibinfo {title} {Non-abelian topological superconductivity in maximally twisted double-layer spin-triplet valley-singlet superconductors},\ }\href {https://doi.org/10.1038/s42005-023-01165-5} {\bibfield  {journal} {\bibinfo  {journal} {Communications Physics}\ }\textbf {\bibinfo {volume} {6}},\ \bibinfo {pages} {47} (\bibinfo {year} {2023})}\BibitemShut {NoStop}%
\bibitem [{\citenamefont {Liu}\ \emph {et~al.}(2023{\natexlab{b}})\citenamefont {Liu}, \citenamefont {Zhou}, \citenamefont {Zhang}, \citenamefont {Chen},\ and\ \citenamefont {Yang}}]{Fan2023}%
  \BibitemOpen
  \bibfield  {author} {\bibinfo {author} {\bibfnamefont {Y.-B.}\ \bibnamefont {Liu}}, \bibinfo {author} {\bibfnamefont {J.}~\bibnamefont {Zhou}}, \bibinfo {author} {\bibfnamefont {Y.}~\bibnamefont {Zhang}}, \bibinfo {author} {\bibfnamefont {W.-Q.}\ \bibnamefont {Chen}},\ and\ \bibinfo {author} {\bibfnamefont {F.}~\bibnamefont {Yang}},\ }\bibfield  {title} {\bibinfo {title} {Making chiral topological superconductors from nontopological superconductors through large angle twists},\ }\href {https://doi.org/10.1103/PhysRevB.108.064508} {\bibfield  {journal} {\bibinfo  {journal} {Phys. Rev. B}\ }\textbf {\bibinfo {volume} {108}},\ \bibinfo {pages} {064508} (\bibinfo {year} {2023}{\natexlab{b}})}\BibitemShut {NoStop}%
\bibitem [{\citenamefont {Lin}\ \emph {et~al.}(2023)\citenamefont {Lin}, \citenamefont {Huang},\ and\ \citenamefont {Lu}}]{Lin_2023}%
  \BibitemOpen
  \bibfield  {author} {\bibinfo {author} {\bibfnamefont {C.}~\bibnamefont {Lin}}, \bibinfo {author} {\bibfnamefont {C.}~\bibnamefont {Huang}},\ and\ \bibinfo {author} {\bibfnamefont {X.}~\bibnamefont {Lu}},\ }\bibfield  {title} {\bibinfo {title} {Customizing topological phases in the twisted bilayer superconductors with even-parity pairings},\ }\href {https://doi.org/10.1088/1674-1056/acd3e3} {\bibfield  {journal} {\bibinfo  {journal} {Chinese Physics B}\ }\textbf {\bibinfo {volume} {32}},\ \bibinfo {pages} {087401} (\bibinfo {year} {2023})}\BibitemShut {NoStop}%
\bibitem [{\citenamefont {Brosco}\ \emph {et~al.}(2024)\citenamefont {Brosco}, \citenamefont {Serpico}, \citenamefont {Vinokur}, \citenamefont {Poccia},\ and\ \citenamefont {Vool}}]{Brosco2024}%
  \BibitemOpen
  \bibfield  {author} {\bibinfo {author} {\bibfnamefont {V.}~\bibnamefont {Brosco}}, \bibinfo {author} {\bibfnamefont {G.}~\bibnamefont {Serpico}}, \bibinfo {author} {\bibfnamefont {V.}~\bibnamefont {Vinokur}}, \bibinfo {author} {\bibfnamefont {N.}~\bibnamefont {Poccia}},\ and\ \bibinfo {author} {\bibfnamefont {U.}~\bibnamefont {Vool}},\ }\bibfield  {title} {\bibinfo {title} {Superconducting qubit based on twisted cuprate van der waals heterostructures},\ }\href {https://doi.org/10.1103/PhysRevLett.132.017003} {\bibfield  {journal} {\bibinfo  {journal} {Phys. Rev. Lett.}\ }\textbf {\bibinfo {volume} {132}},\ \bibinfo {pages} {017003} (\bibinfo {year} {2024})}\BibitemShut {NoStop}%
\bibitem [{\citenamefont {Patel}\ \emph {et~al.}(2024)\citenamefont {Patel}, \citenamefont {Pathak}, \citenamefont {Can}, \citenamefont {Potter},\ and\ \citenamefont {Franz}}]{Patel2024}%
  \BibitemOpen
  \bibfield  {author} {\bibinfo {author} {\bibfnamefont {H.}~\bibnamefont {Patel}}, \bibinfo {author} {\bibfnamefont {V.}~\bibnamefont {Pathak}}, \bibinfo {author} {\bibfnamefont {O.}~\bibnamefont {Can}}, \bibinfo {author} {\bibfnamefont {A.~C.}\ \bibnamefont {Potter}},\ and\ \bibinfo {author} {\bibfnamefont {M.}~\bibnamefont {Franz}},\ }\bibfield  {title} {\bibinfo {title} {$d$-mon: A transmon with strong anharmonicity based on planar $c$-axis tunneling junction between $d$-wave and $s$-wave superconductors},\ }\href {https://doi.org/10.1103/PhysRevLett.132.017002} {\bibfield  {journal} {\bibinfo  {journal} {Phys. Rev. Lett.}\ }\textbf {\bibinfo {volume} {132}},\ \bibinfo {pages} {017002} (\bibinfo {year} {2024})}\BibitemShut {NoStop}%
\end{thebibliography}%

~
\newpage
\pagebreak

\appendix

\setcounter{figure}{0}
\setcounter{table}{0}
\makeatletter
\renewcommand{\thefigure}{S\arabic{figure}}

\section{Edge current from spectral function}

In this Section we give details pertaining to the derivation of Eq.\ \eqref{scurr}. We begin by rewriting Eq.\ \eqref{curr1} for the bond current operator as
\begin{align}\label{eq:curr2}
j_{ij}=&-it\left(c^\dagger_{i\uparrow}c_{j\uparrow}+c_{i\downarrow}c^\dagger_{j\downarrow}\right) +{\rm h.c.} \nonumber \\
=&-it {\rm Tr}(\psi_j\psi_i^\dagger)+{\rm h.c.} ,    
\end{align}
where we defined a Nambu spinor $\psi_j=(c_{j\uparrow},c^\dagger_{j\downarrow})^T$. To avoid clutter we focus here on the current in a single layer and take $e=\hbar=1$. Regarding $\psi_j$ as a field operator in imaginary time we can further express the current expectation value as
\begin{align}\label{eq:curr3}
\langle j_{ij}\rangle=&-it\lim_{\tau\to 0^+}{\rm Tr}\langle\psi_j(\tau)\psi_i^\dagger(0)\rangle +{\rm c.c.} \nonumber \\ =& -it {\rm Tr}\, \cG_{ji}(\tau=0^+)+{\rm c.c.},    
\end{align}
where $\cG_{ji}(\tau)=\langle T_\tau \psi_j(\tau)\psi_i^\dagger(0)\rangle$ denotes the imaginary time Gorkov Green's function.

For a long strip geometry it is convenient to switch to a notation there $\br=(x_j,y_j)$ denotes the position of site $j$. The current along an $x$-bond distance $y$ from the edge can then be expressed as 
\begin{equation}\label{eq:curr4}
    J_{\hat{x}}(y)=-it{1\over N_x}\sum_x{\rm Tr}\,\cG_{\br+\hat{x},\br}(0^+)+{\rm c.c.}.
\end{equation}
Exploiting translation invariance along $x$ we introduce mixed Fourier representation $\cG_{\br+\hat{x},\br}(\tau)=N_x^{-1}\sum_k e^{ik}\cG_k(y,\tau)$ and obtain for the current 
\begin{equation}\label{eq:curr5}
    J_{\hat{x}}(y)=2t{1\over N_x}\sum_k \sin{k}\ {\rm Tr}\, \cG_{k}(y,0^+).
\end{equation}
As the final step we pass to the Matsubara frequency $\cG_{k}(y,\tau)=\beta^{-1}\sum_n e^{-i\omega_n\tau}\cG_k(y,\omega_n)$ and express $\cG$ in terms of its spectral representation 
\begin{equation}\label{eq:cG}
\cG_k(y,\omega_n)=\int_{-\infty}^\infty {d\omega'\over 2\pi}{A_k(y,\omega')\over i\omega_n-\omega'}
\end{equation}
with $A_k(y,\omega)=-2{\rm Im}G^{\rm ret}_k(y,\omega)$. Here $G^{\rm ret}$ denotes the retarded Green's function obtained from $\cG$ by analytic continuation $i\omega_n\to \omega+i\delta$. Substituting Eq.\ \eqref{eq:cG} to \eqref{eq:curr5} and performing the required Matsubara sum using 
\begin{equation}\label{eq:mats}
{1\over \beta}\sum_n {e^{-i\omega_n0^+}\over i\omega_n-\omega}={1\over e^{\beta\omega}+1}
\end{equation}
we obtain Eq.\ \eqref{scurr} of the main text.

\section{Maximal gap at Fermi surface for $d-$wave superconductors}\label{app:max_gap}

Our goal is to find an approximate maximum value of the SC order parameter at the Fermi surface for both layers. Let us assume decoupled layers for now. Layer 1 has the $d_{x^2-y^2}$ order parameter and layer 2 has $d_{xy}$ order parameter. Regularized on a square lattice, the superconducting order parameters for layers 1 and 2 are
\begin{align}
    \Tilde{\Delta}_1(\mathbf{k})&=\Delta_1[\text{cos}(k_xa)-\text{cos}(k_ya)]\\
    \Tilde{\Delta}_2(\mathbf{k})&=\Delta_2[2\text{sin}(k_xa)\text{sin}(k_ya)].  
\end{align}
By symmetry, we know that the nodes of $\Tilde{\Delta}_1(\mathbf{k})$ correspond to the maxima of $\Tilde{\Delta}_2(\mathbf{k})$ and vice versa. This gives us the conditions for maximizing the gap:
\begin{itemize}
    \item $\Tilde{\Delta}_1(\mathbf{k})\vert_{\text{max}}$ occurs when $k_x.k_y=0$ (nodes of $\Tilde{\Delta}_2(\mathbf{k})$).
    \item $\Tilde{\Delta}_2(\mathbf{k})\vert_{\text{max}}$ occurs when $|k_x|=|k_y|$ (nodes of $\Tilde{\Delta}_1(\mathbf{k})$).
\end{itemize}
However, we need to maximize the gap on the Fermi surface. The kinetic energy and the on-site potential of the Hamiltonian are
\begin{equation}
    \xi(\mathbf{k})=-2t[\cos(k_xa)+\cos(k_ya)]-\mu.
\end{equation}
We define the Fermi surface when $\xi(\mathbf{k})=0$. Let $A=\frac{-\mu}{2t}$. Then, at the Fermi surface we have
\begin{equation}
    \cos(k_xa)+\cos(k_ya)=A.
\end{equation}
Given the constraint imposed by the Fermi surface above, we can now find the maximum expected gap for both layers respectively.\\ %

\noindent For layer 1, we maximize the gap ($\Tilde{\Delta}_{1,\text{max}}$) using the following set of equations:
\begin{align}
    \cos(k_xa)+\cos(k_ya)&=A\\
    \cos(k_xa)-\cos(k_ya)&=\frac{\Tilde{\Delta}_{1,\text{max}}}{\Delta_1}\\
    k_x.k_y&=0
\end{align}
Assuming $k_x=0$ as a solution, we get the form of the maximum gap as
\begin{equation}\label{d1max}
    \Tilde{\Delta}_{1,\text{max}}=2\Delta_1\left(1+\frac{\mu}{4t}\right).
\end{equation}

\noindent For layer 2, we maximize the gap ($\Tilde{\Delta}_{2,\text{max}}$) using the following set of equations:
\begin{align}
    \cos(k_xa)+\cos(k_ya)&=A\\
    2.\sin(k_xa).\sin(k_ya)&=\frac{\Tilde{\Delta}_{2,\text{max}}}{\Delta_2}\\
    |k_x|=|k_y|
\end{align}
Assuming $k_x=k_y$ as a solution, we get the form of the maximum gap as
\begin{equation}\label{d2max}
    \Tilde{\Delta}_{2,\text{max}}=2\Delta_2\left[1-\left(\frac{\mu}{4t}\right)^2\right].
\end{equation}



\end{document}